\documentclass[aps,prb,twocolumn,showkeys]{revtex4-1}
\usepackage{graphics,graphicx,epsfig,color,hyperref,latexsym,float,amsmath}
\usepackage{times}
\input pdfcolor.tex

\begin{document}
\title{
Thermal decoherence in a strongly correlated Bose liquid
}
\author{Abhishek Joshi and Pinaki Majumdar}
\affiliation{Harish-Chandra Research Institute, HBNI, Chhatnag Road, Jhusi,
Allahabad 211019}
\date{\today}

\begin{abstract}
We compute the single particle spectral function of a Bose liquid on a 
lattice, at integer filling, close to the superfluid-Mott transition. 
We use a `static path approximation'  that retains all the classical 
thermal fluctuations in the problem, and a real space implementation of 
the random phase approximation (RPA) for the Green's functions on the 
thermally fluctuating backgrounds. This leads to the standard RPA answers 
in the ground state but captures the progressive damping of the excitations 
with increasing temperature. We focus on the momentum resolved lineshape
across the superfluid to Bose liquid thermal transition. In the superfluid 
regime we observe gapped `amplitude' modes, and gapless `phase' modes of 
positive and negative energy. The dispersion and weight of these modes 
changes with interaction but are almost temperature independent, even into
the normal state, except near critical coupling. The damping of the modes
varies roughly as $T^{\alpha} f_{\bf k}$, where $T$ is the temperature and 
${\bf k}$ the momentum, with $\alpha \sim 0.5$ and $f_{\bf k}$ being weakly
momentum dependent. The Mott phase has gapped dispersive spectra. 
Near critical coupling the thermal Bose `liquid' is gapped, with progressive 
widening of the gap with increasing temperature, a feature that it shares 
with the Mott insulator.
\end{abstract}

\maketitle

\section{Introduction}

The homogeneous Bose gas with weak repulsive interactions 
\cite{nozieres-qlig} has a dispersion 
 $\omega_{\bf k} = v_s k$, where ${\bf k}$ is the
momentum of the excitation, $k = \vert {\bf k} \vert$, and
$v_s$ is the critical superfluid (SF)
velocity. This result differs from the
$\omega_{\bf k} \propto k^2$ that one expects for free particles,
and is responsible for supporting `superflow' upto a velocity
$\sim v_s$ in the system. The result was arrived at by using
a method suggested by Bogoliubov \cite{bog-ref}.
In the presence of a potential (a trap, say) the order parameter
profile is described by the Gross-Pitaevskii 
equation \cite{gp-eqn}, 
and excitations are obtained by using a Gaussian
expansion \cite{stringari-rmp} on this background.

The presence of a lattice, as in ultracold atomic 
systems \cite{ultracold-genl},
has two effects, (i)~it affords 
tunability of the interaction to kinetic energy 
ratio (for a given atomic species), and 
(ii)~it breaks galilean invariance.
At integer density, increasing the lattice depth drives 
a transition from the superfluid to a Mott insulator 
\cite{jaksch,bloch-rev}.
The weak coupling superfluid, as in the continuum, has a 
low energy
linearly dispersing mode, while the Mott insulator - a phase 
specific to the lattice - has a gapped excitation spectrum. 
Increasing interaction in the superfluid leads to the appearance of a
`negative energy' gapless mode - which dictates the broadening of the
momentum distribution from the simple ${\bf k} =0$ peak, and gapped
`amplitude' modes of both positive and negative energy
As the system heads towards the Mott transition the weight
shifts from the traditional Bogoliubov mode to the negative energy phase
mode and the amplitude modes.

The realization of a superfluid to Mott transition
\cite{Bloch, Immanuel Bloch, Zwerger} in optical lattices
has allowed access to some of the dynamical properties of
correlated bosons.
The excitation spectra can be probed via 
Bragg spectroscopy \cite{Fort, Ernst, Klaus},
which accesses the dynamic structure factor,  
and lattice spectroscopy \cite{Endres}, 
which probes the kinetic energy correlator.
The measurements have revealed the existence of  
of the sound (Bogoliubov) mode and also the 
more  exotic amplitude mode.

\begin{figure}[b]
\centerline
{
\includegraphics[width=3.5cm,height=4.0cm]{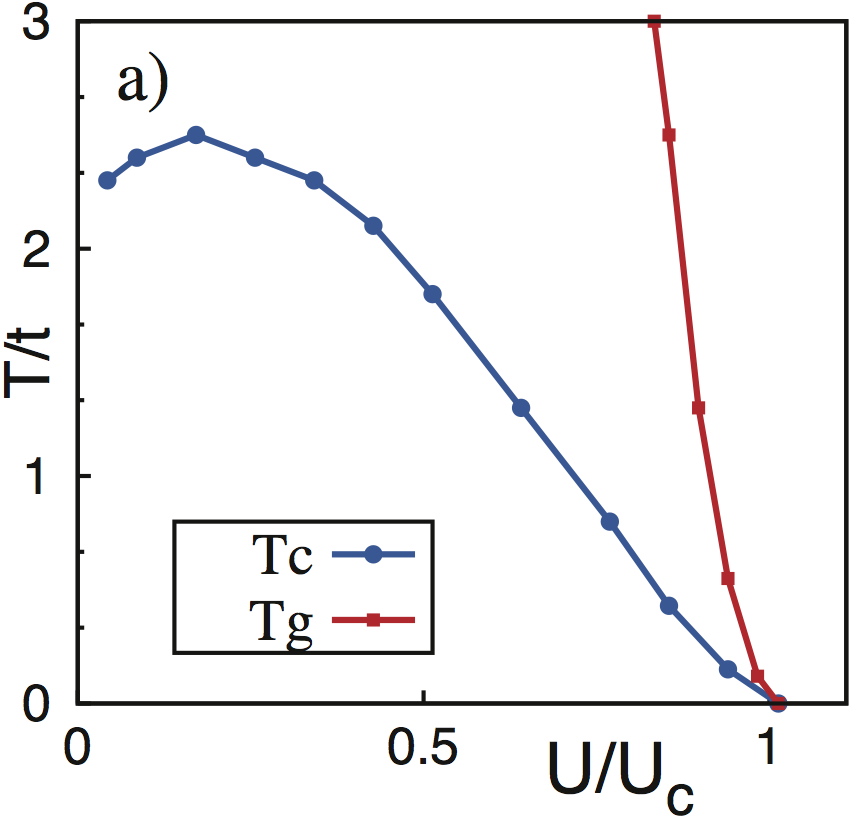}
\includegraphics[width=4.5cm,height=4.1cm]{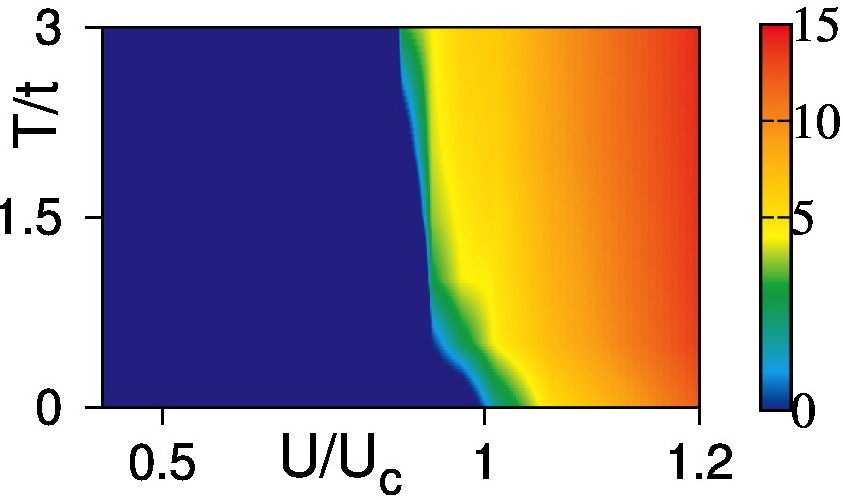}
}
\caption{Behaviour of the low energy spectral gap for
varying interaction, $U/U_c$, where $U_c$ is the coupling 
for SF-Mott transition,  and temperature.
(a)~Shows the SF to normal transition temperature, $T_c$, and the temperature
$T_g$ above which the spectrum shows a gap. One expects a gap for $U > U_c$, 
but we find that already for $U/U_c \sim 0.9$ the `normal Bose liquid' is 
gapped at high temperature. 
(b)~Magnitude of the gap for 
varying $U$ and $T$. Blue indicates ungapped and we emphasize the
small gapped region for $U/U_c < 1$.
}
\end{figure}

There have been efforts in
calculating the single particle and 
two particle spectral properties.
At zero temperature, the methods employed include 
a strong coupling approach
\cite{Sen}, Schwinger-boson mean field theory \cite{Huber}, 
the random phase approximation \cite{Menotti}, the quantum
rotor approach\cite{TA1} 
and a variational cluster method \cite{Knap1,Knap2,Knap3,Knap4}.
Their main result is that the superfluid 
spectrum consists of two gapless modes  and a
gapped mode. The weight in  the gapped mode increases as one 
moves toward the transition, while the `gap' itself reduces.
There have also been attempts to go beyond the RPA formalism  
within the Mott phase\cite{Blatter,Pelster}.

The effect of thermal fluctuations 
on the single particle spectrum 
seem to have seen much less of an effort.
This is a fascinating question in the vicinity of
a Mott transition since results on the fermionic
problem indicate that the coherence temperature of 
a Fermi liquid vanishes 
\cite{georges-krauth-prb,dmft-rmp}
as it approaches the Mott transition.
For bosons in one dimension quantum Monte Carlo (QMC)
results \cite{Peter} indicate strong broadening of the 
particle band 
and suppressed spectral weight for hole excitations. The 
broadened positive branch has no clear linear behaviour 
at small momentum. Recent QMC studies 
in higher dimension are more focused on calculating
two particle response functions\cite{Liu,Kun,Pollet}.
Other approaches  to thermal properties 
are slave bosons \cite{stoof}, the quantum 
rotor approach \cite{TA2} 
in 2D and 3D and bosonic 
dynamical mean field theory \cite{Panas} 
(BDMFT) in 3D.
The slave rotor approach indicates an increase in the 
gap with rising temperature with the tip of the Mott 
lobe moving to 
smaller interaction.

We provide an alternate approach to compute single 
particle quantities via a two step process: (i)~we 
`solve' the Bose Hubbard model 
via an auxiliary field decomposition of the
kinetic term and a static (but spatially fluctuating)
approximation for the auxiliary field, and
 (ii)~on the
equilibrium configurations of this theory, which capture $T_c$
quite well \cite{bhm-thermal}, 
we use a real space generalisation of the 
`random phase approximation' (RPA) for the boson Green's
function. 
The method has no analytic continuation problems
and directly computes real frequency correlation functions
as we will explain later.
Both (i) and (ii) above readily generalise 
to the presence of a trap, background
disorder, and effects like spin-orbit coupling.
Our main results are the following:
\begin{enumerate}
\item
{\it Nature of the spectral function:}
At a general coupling, $U$, and temperature $T$, 
the spectral weight in the 
single particle Green's function in the superfluid 
resides primarily in two gapless `phase modes' and
two gapped `amplitude modes', each with a positive and negative
energy branch. The lineshape $A({\bf k}, \omega)$ 
can be characterised by a `four peak' structure indexed 
by their mean excitation energy $\Omega_{n,{\bf k}}$, 
the weight (or residue) $r_{n,{\bf k}}$,
and the width (or damping) $\Gamma_{n,{\bf k}}$. The
integer $n$ indexes the four peaks at a given momentum.
\item
{\it Temperature dependence:} 
While the $\Omega,~r,~\Gamma$ all depend on $U$,
the $\Omega_{n,{\bf k}}$ and  $r_{n,{\bf k}}$ are
essentially temperature independent in the superfluid, 
except near $U_c$.  $\Gamma_{n,{\bf k}}$, within our 
scheme, is zero at $T=0$ and can be 
approximated by $\Gamma_{n,{\bf k}} 
\sim T^{\alpha} f_{n,{\bf k}}$,
with $\alpha \sim 0.5$ and the 
$f_{n,{\bf k}}$ are weakly momentum dependent.
\item
{\it Vicinity of the Mott transition:}
In the vicinity of critical coupling,
$U \gtrsim 0.9 U_c$,  the normal `Bose liquid' is gapped, 
with progressive widening of the gap with increasing 
temperature, Fig.1.
This is a feature that it shares 
with the finite temperature Mott insulator.
\item
{\it Interaction effects in the normal Bose liquid:}
In the high temperature normal state 
a two band description suffices. The dispersion is gapped
once beyond an interaction $U > U_{gap}(T)$. 
While the bandwidth increases monotonically as $U$ 
increases towards $U_c$, the 
damping, roughly momentum independent,
peaks at a scale $U \sim 4t-5t$ 
(around where $T_c$ also
peaks) and reduces as $U \rightarrow U_c$. 
\end{enumerate}

The paper is organized as follows: in 
Section II we discuss the model, our method, and its 
numerical implementation. 
Section III discusses our results for the ground
state, Section IV discusses thermal effects in different
interaction regimes, while Section V highlights the effect
of increasing interaction on the high temperature
normal Bose liquid. 
We conclude in Section VI.

\newpage

\section{Model and method}

The Bose Hubbard model (BHM) is given by.
\begin{equation}
H = -t\sum_{<ij>}(b^{\dagger}_i b_j ~+~ h.c.)
- \mu \sum_i n_i
+ (U/2)  \sum_{i} n_i(n_i-1) \nonumber
\end{equation}
where $t$ is the nearest neighbour 
hopping, $U$ the  interaction strength and
 $\mu$ the chemical potential. In an earlier paper \cite{bhm-thermal}
we have discussed in detail the method for obtaining
the thermal phase diagram of the model above by
using an auxiliary field decomposition of the
kinetic term and then either (i)~a classical
(time independent) approximation for the auxiliary
field $\psi_i(\tau)$ 
- called the static path approximation (SPA), 
or (ii)~retaining Gaussian quantum fluctuations about
the SPA result, the so called perturbed SPA (or PSPA)
scheme. We quickly recapitulate the SPA scheme below.
Once the equilibrium distribution for the auxiliary
field is obtained, via a Monte Carlo strategy, we 
can compute the boson Green's function through a
real space implementation of the random phase
approximation (RPA). The RPA, although not
exact, has been shown to yield qualitatively correct
results in the ground state.

The BHM in path integral formalism can 
be written as 
\begin{eqnarray}
Z &=& 
\int {\cal D} \phi {\cal D} \phi^* e^{-(S_0+S_K)}\nonumber \cr 
S_0 &=& \int_0^{\beta} d\tau 
[\sum_i
{\phi^*_i} (\partial_{\tau} - \mu) \phi_i 
+ {U \over 2} \sum_i \phi^*_i \phi_i(\phi^*_i \phi_i -1)] \cr 
S_K &=& \int_0^{\beta} d\tau (-t)\sum\limits_{<ij>}
(\phi^*_i\phi_j + h.c.) \nonumber
\end{eqnarray}
where $\phi$ represents the original bosonic fields.

This can be rewritten as 
\begin{eqnarray}
Z~ &=& 
\int 
{\cal D} \phi {\cal D} \phi^* {\cal D} \psi {\cal D} \psi^* 
\text{e}^{-(S+S_b)} \cr
S~ & =& S_0[\phi] 
-\int_0^\beta d\tau\sum_{ij}(C_{ij}
{\phi}_{{i}}^*(\tau)\psi_{{j}}+hc) +\beta\sum_{i}{\psi}_{i}^*\psi_{i}\cr
C_{ij} & =& 
\frac{1}{N}
\sum\limits_{\vec{k}}\sqrt{A_{\vec{k}}}
e^{i\vec{k}(\vec{r_i}-\vec{r_j})},~~
B_{ij}  = 
\frac{1}{N}
\sum\limits_{\vec{k}}\sqrt{B_{\vec{k}}}
e^{i\vec{k}(\vec{r_i}-\vec{r_j})} \cr
S_b~&=&-\sum\limits_{ij}
B_{ij}{\phi^*_{i,0}}\phi_{j,0}-
\sum_{<i,j>}^{n \neq 0} t_{ij}
{\phi^*_{i,n}}\phi_{j,n}\nonumber
\end{eqnarray}
with $ A_{\vec{k}}  = \theta( t_{\vec{k}}) 
t_{\vec{k}} $ and
$B_{\vec{k}}  = \theta(-t_{\vec{k}})  t_{\vec{k}}$\\
\newline where $t_{\vec{k}}=2 t (cos(k_x a)+cos(k_y a))$

We have to compute 
\begin{eqnarray}
G_{ij}(i\omega_n) &=&  Tr[e^{-\beta H} b_{j,n} 
b^{\dagger}_{i,n}] \cr
&=& ~\int 
{\cal D} \phi {\cal D} \phi^* {\cal D} \psi {\cal D} \psi^*
\text{e}^{-(S+S_b)} \phi_{j,n} \phi^{*}_{i,n}\cr
&=& ~\int 
{\cal D} \phi {\cal D} \phi^* {\cal D} \psi {\cal D} \psi^*
\text{e}^{-S}F(\phi_{j,n},\phi_{i,n}^*)\cr
F(\phi_{j,n},\phi_{i,n}^*)&=&\phi_{j,n} \phi^{*}_{i,n}(1-S_b+\frac{S_b ^2}{2!}+...)
\nonumber\cr
\end{eqnarray}

\begin{figure*}[t]
\centerline
{
~~~~~~
\includegraphics[width=13.0cm,height=4.6cm]{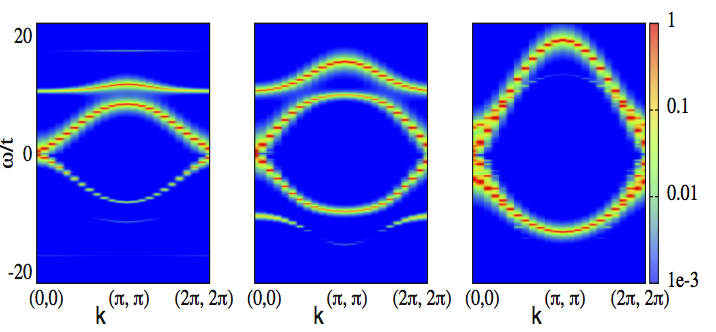}
}
\vspace{.2cm}
\centerline
{
\includegraphics[width=4.1cm,height=3.2cm]{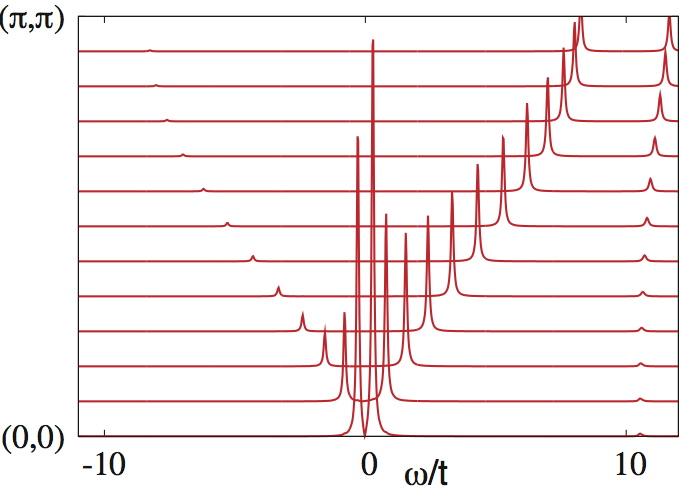}
\includegraphics[width=4.1cm,height=3.2cm]{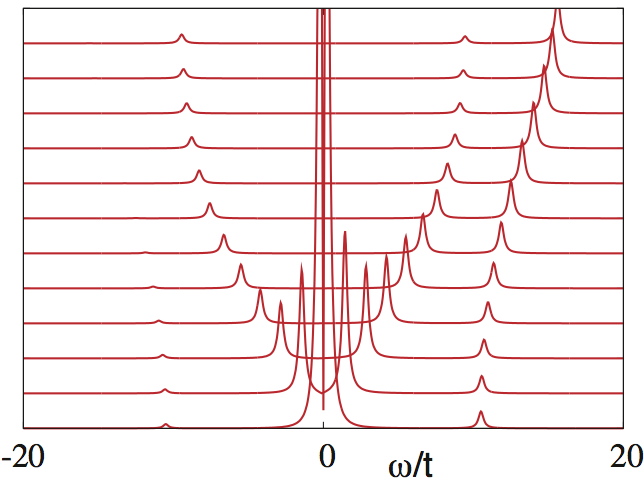}
\includegraphics[width=4.1cm,height=3.2cm]{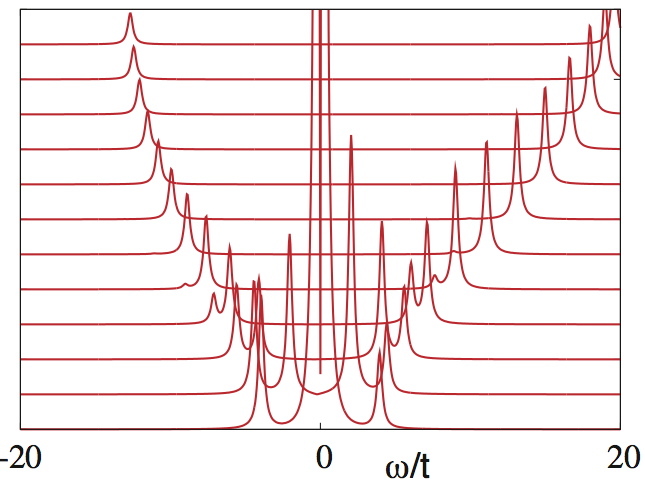}
~~
}
\vspace{.2cm}
\centerline
{
\includegraphics[width=4.0cm,height=3.2cm]{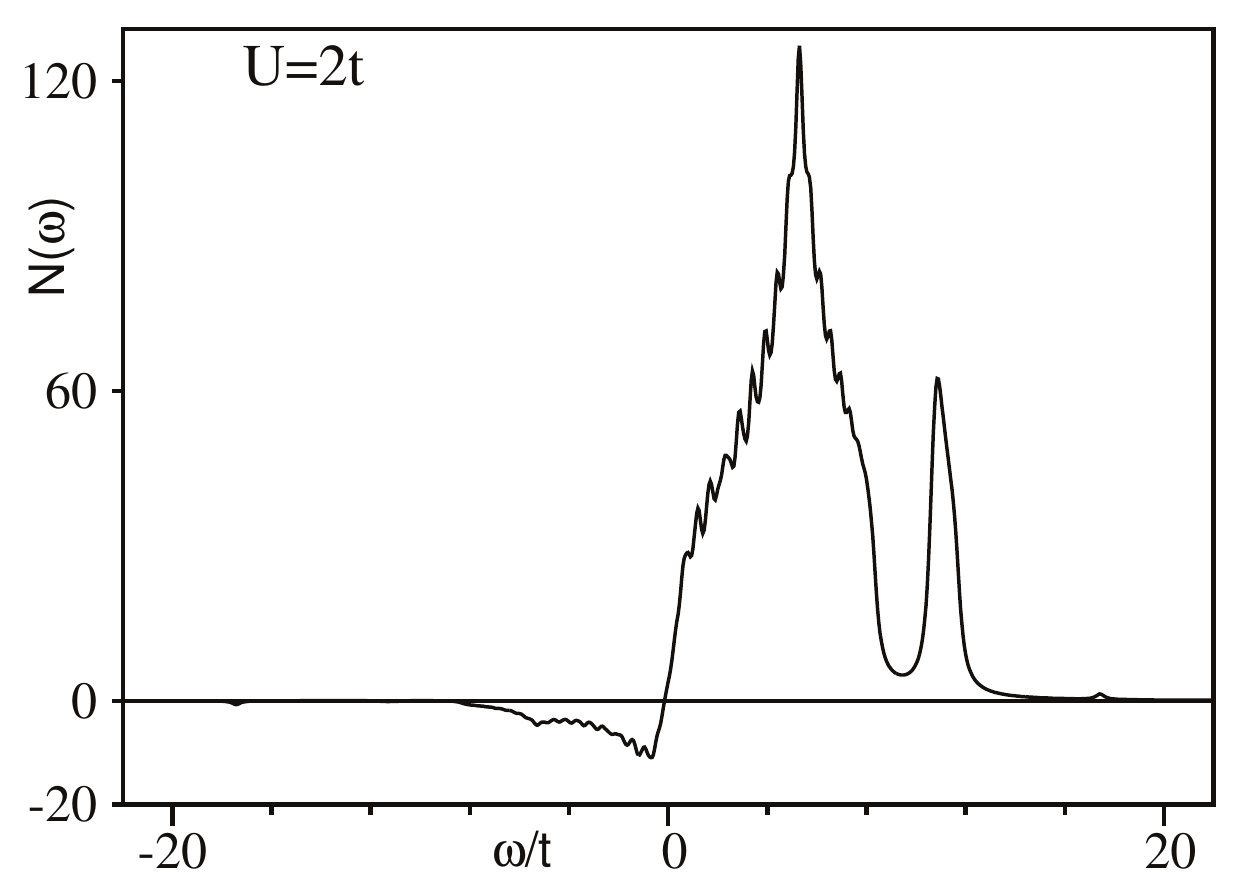}
\includegraphics[width=4.0cm,height=3.2cm]{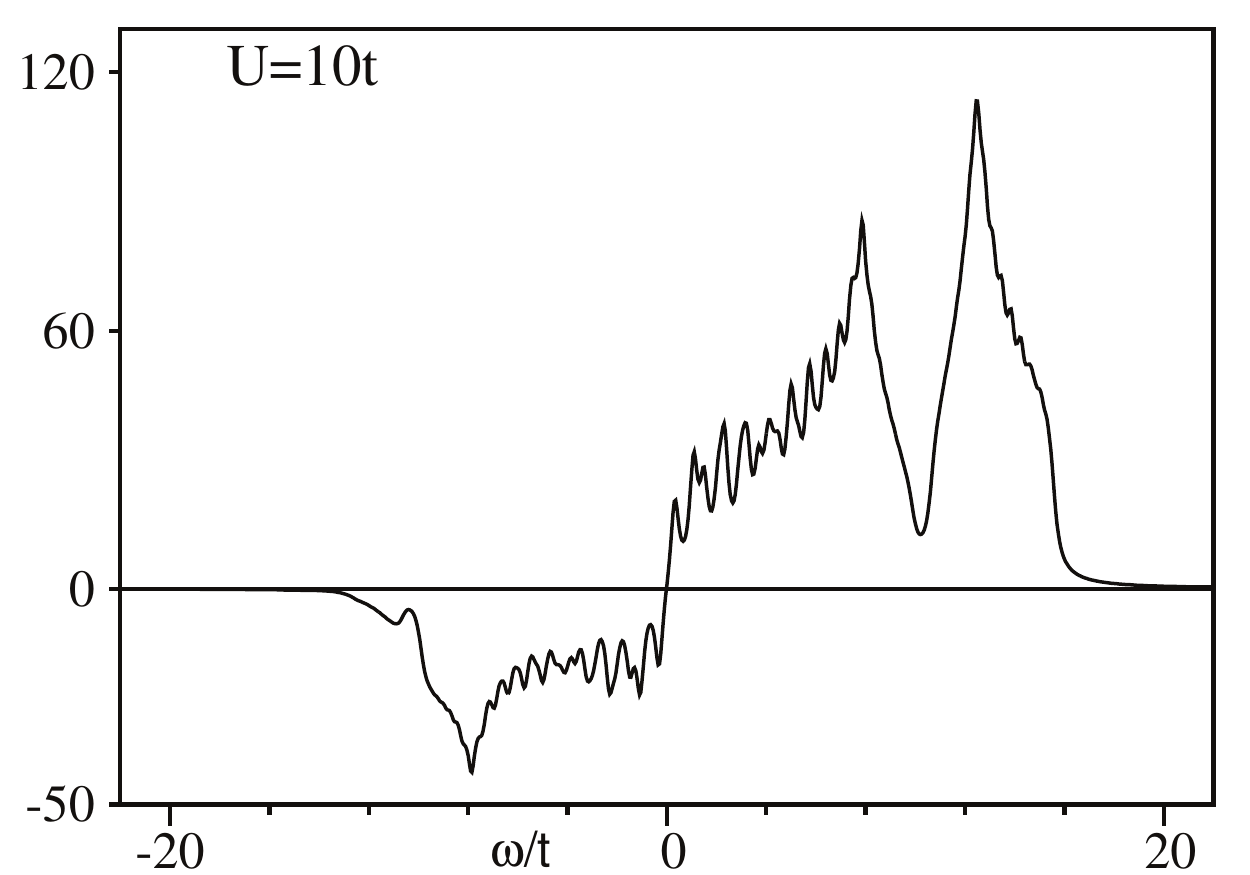}
\includegraphics[width=4.0cm,height=3.2cm]{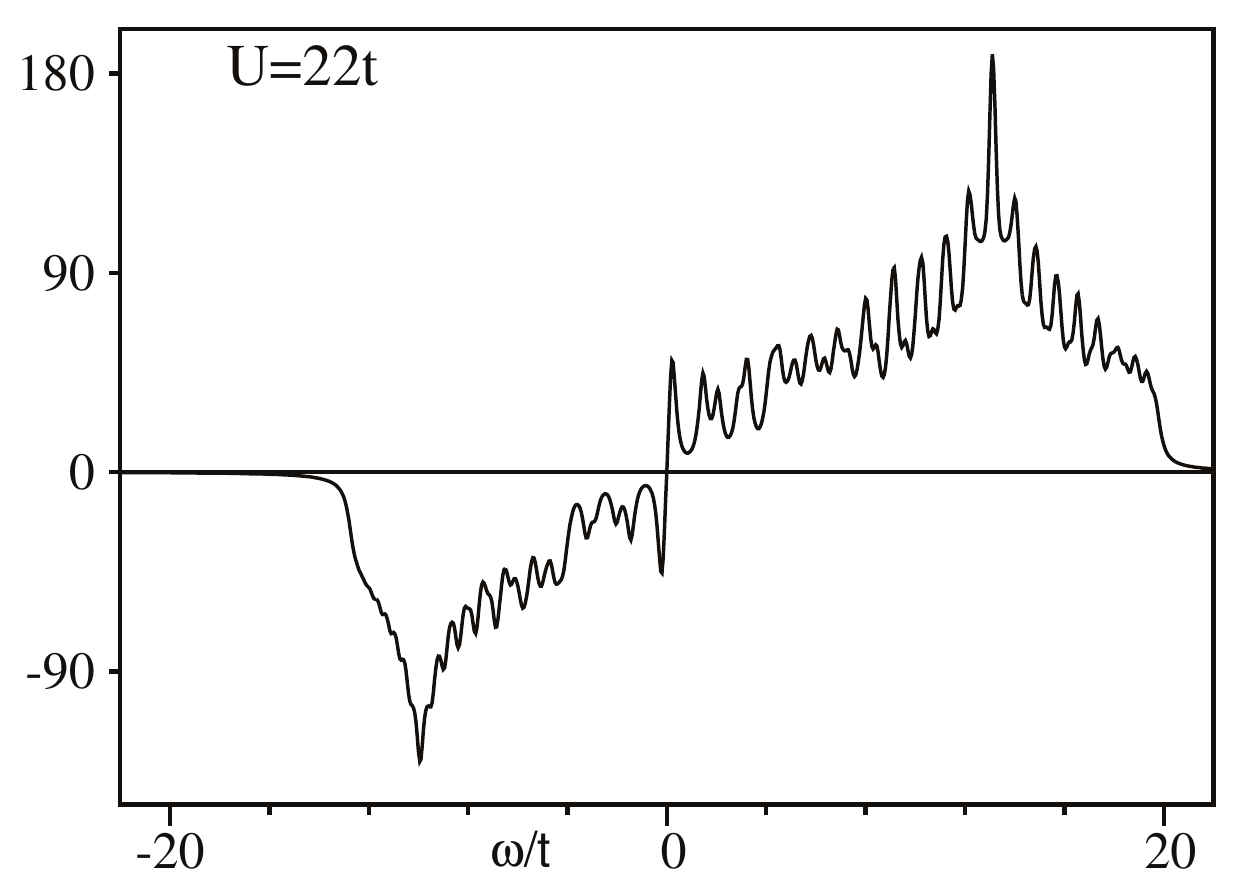}
}
\caption{Spectral function and density of states in the
superfluid ground state. From left to right along each row,
 $U = 2t,~10t,~22t$. First row:
$|A({\bf k}, \omega)|$ second row: the lineshape of $\vert 
A({\bf k}, \omega) \vert$,
and  third row: the density of states $N(\omega)$.}  
\end{figure*}

We compute the Green's function via the
`self avoiding walk' method proposed in the  context of the
strong coupling expansion. On the mean field state it leads to
the same answers as standard RPA but there are no answers
yet on the finite temperature backgrounds. The detailed
method is discussed in the Appendix, we only quote the
final answer here.
$$
\hat{G}=\int {\cal D}\psi 
{ \cal D}\psi^*e^{-\sum_i\psi^*_i\psi_i} \hat{{\cal G}}
$$ 
${\cal \hat{G}}=
\frac{ \hat{g}}{(1-{\cal{T}}\hat{g})}=\frac{I}{(\hat{g}^{-1}-{\cal{T}})}$
where $\hat{g}$ is the Green's 
function of the atomic problem and
$\cal{T}$ is the hopping matrix.
We analytically continue 
$G(i\omega_n\rightarrow \omega+i\eta)$
to obtain the retarded Green function.
This is the RPA result averaged over thermal
 configurations  of the $\psi$ field.
 Now,  $A({\bf k}, \omega)=-\int {\cal D}\psi 
{ \cal D}\psi^*e^{-\sum_i\psi^*_i\psi_i} 
\frac{1}{\pi}Im(\hat{v}\hat{\cal{G}}(\omega)\hat{v}\dagger)$
and $N(\omega)={1 \over N} \sum_{\bf k} A({\bf k},\omega)$.

We solve the SPA hamiltonian for each background configuration.
Then we construct atomic Green's function matrix $\hat{g}$
whose expressions are given in the Appendix. $\hat{g}$
is the block diagonal matrix of size $2(L*L)\times 2(L*L)$.
$\cal{T}$ is the hopping matrix of size $2(L*L)\times2(L*L)$
whose expressions are also given in the Appendix.
To compute the corrected Green's function for each background
we subtract the hopping matrix from the inverted
the atomic Green's function
and take its inverse for every value of $\omega+i\eta$.
To compute $A({\bf k},\omega)$ we multiply the above matrix 
$\cal{G}$ by row matrix $\hat{v}$ of size $1\times2(L*L)$
and its hermitian conjugate. We repeat the above procedure for each
thermal configuration and take the average.
All our results are on size $24 \times 24$.

Finally, given the observed `four peak' structure of the spectral
function (with higher bands having negligible weight)
we tried a four Lorentzian fit to the spectral data.
These involve twelve parameters, 4 residues, 4
`center frequencies' and 4 broadenings, with the sum of
the 4 residues being $\sim 1$ acting as a sum rule
check. This is a scheme that we applied at all $T$ in
the $U < U_c$ window, while for $U > U_c$ a two
Lorentzian fit is more appropriate.
Specifically, we tried:
$$
A({\bf k}, \omega) = {1 \over \pi}
\sum_n  r_{n,{\bf k}}  { \Gamma_{n,{\bf k}}
\over { (\omega - \Omega_{n,{\bf k}})^2 + \Gamma_{n,{\bf k}}^2 }}
$$
Much of the analysis in this paper is based on the
estimated parameter set $\{ \Omega_{n,{\bf k}},  r_{n,{\bf k}}, 
\Gamma_{n,{\bf k}} \}$ for varying $U$ and $T$.

\section{Results at zero temperature}

All the results we discuss are at fixed $\mu$
such that particle density is fixed to unity.
We first discuss the zero temperature results,
in terms of two regimes: 
(i)~the  superfluid phase, all the 
way from weak coupling to the 
transition point,
and (ii)~the Mott phase.
The finite temperature results are discussed in a 
similar spirit.

\subsection{The superfluid ground state}

Fig.2 shows the evolution of the spectrum 
in the superfluid 
phase from weak interaction, $U=2t$,
 to intermediate, $U=10t$,
 and then to $U= 22t$. The critical coupling is 
$U_c = 24t$ in the problem. 
The first row shows a map of  
$\vert  A({\bf k}, \omega) \vert$, 
for ${\bf k}$ varying from $(0,0) \rightarrow (\pi,\pi) 
\rightarrow (2\pi, 2\pi)$. The
second row shows the lineshape of $\vert  A({\bf k}, \omega) \vert$,
since some features are not clear in the spectral map,
while the third shows the DOS $N(\omega)$.
We note that
$$
 A({\bf k}, \omega) = \theta (\omega) \vert  A({\bf k}, \omega) \vert
 - \theta(-\omega) \vert  A({\bf k}, \omega) \vert
$$
$\vert  A({\bf k}, \omega) \vert$ therefore contains the 
same information as $A({\bf k}, \omega)$.

In the weak coupling superfluid at $T=0$ the only mode that
has substantial weight is the positive energy gapless mode.
This is the Bogoliubov mode associated with the broken symmetry
- the Goldstone mode in this problem.
When the dominant occupancy is still that of ${\bf k}=(0,0)$, 
the negative energy spectra has very small residue
at all ${\bf k} \neq (0,0)$ since
$
\int_{-\infty}^0 d\omega A({\bf k}, \omega) = n_{\bf k}
$
and we assume that $n_{\bf k} \rightarrow 0$ for all ${\bf k}
\neq 0$.

\begin{figure}[b]
\centerline{
\includegraphics[height=3cm,width=4cm]{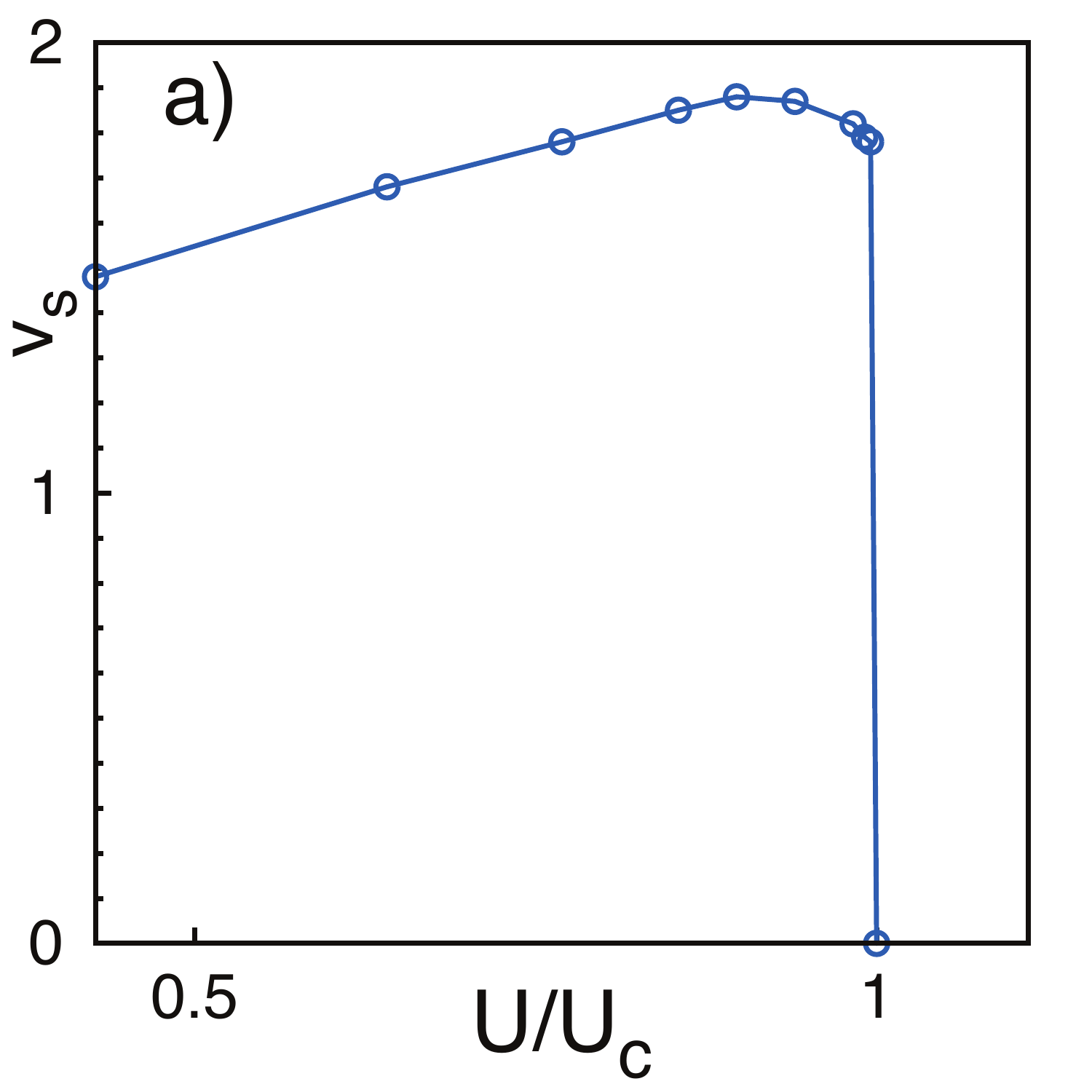}
\includegraphics[height=3cm,width=4cm]{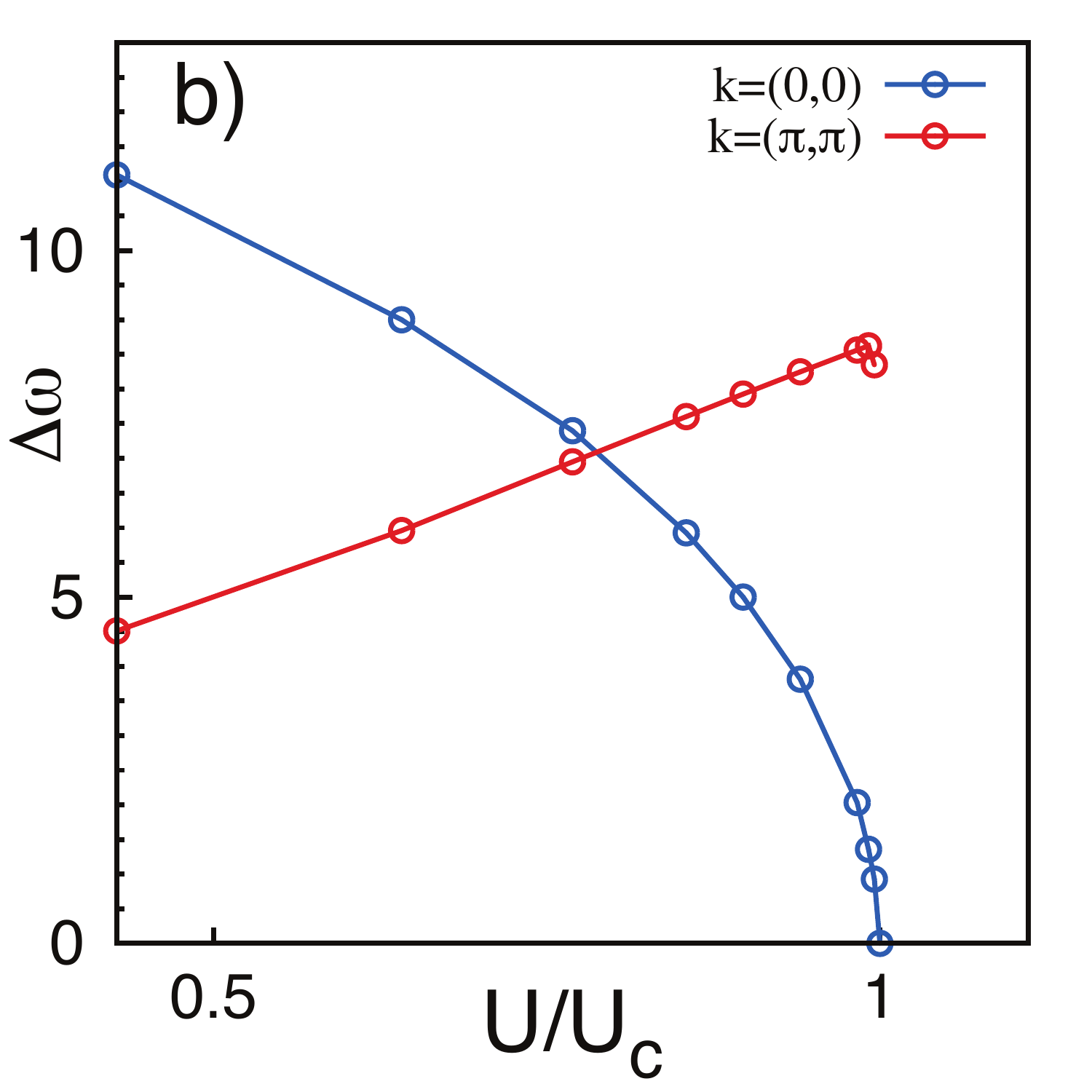}
}
\centerline{
\includegraphics[height=3cm,width=4cm]{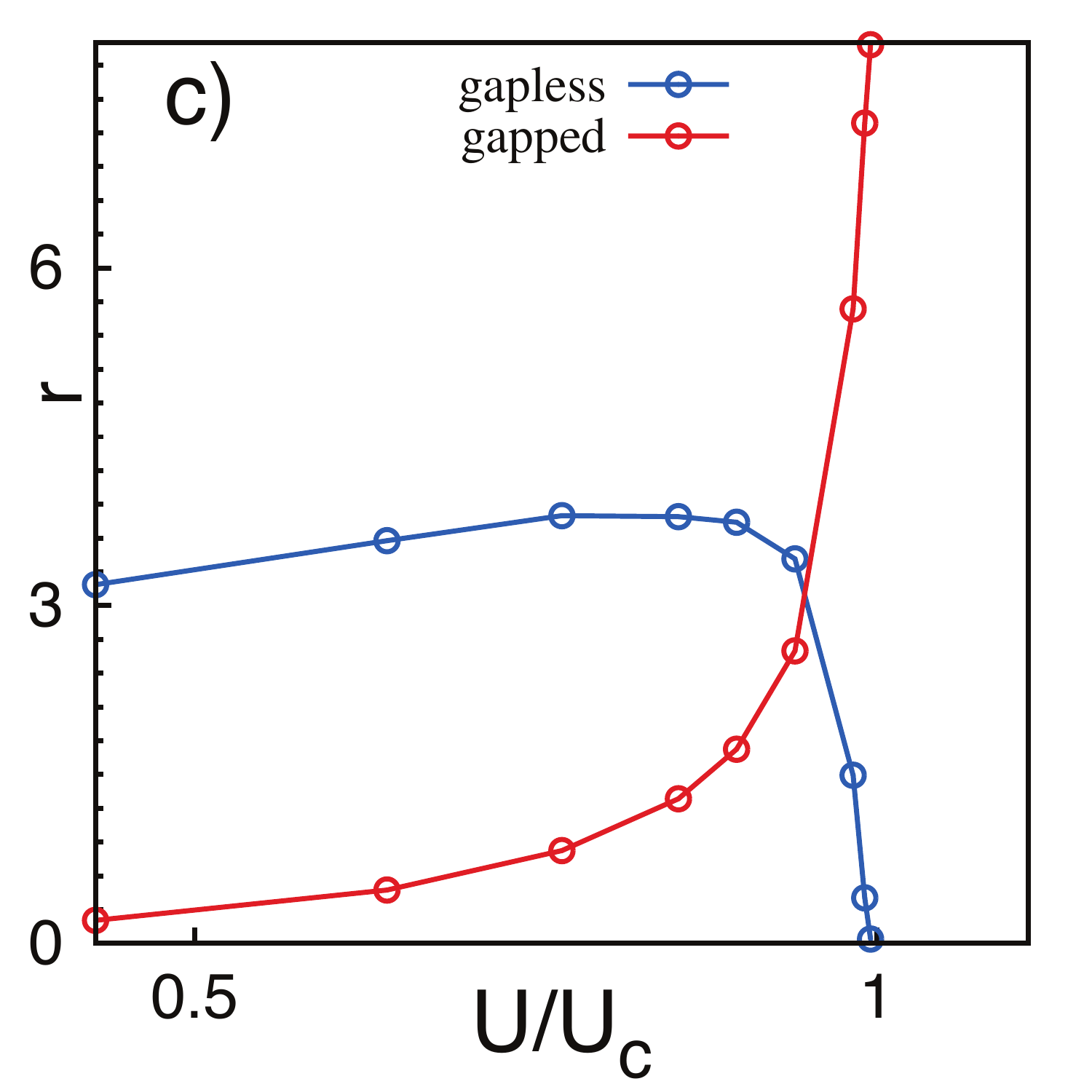}
\includegraphics[height=3cm,width=4cm]{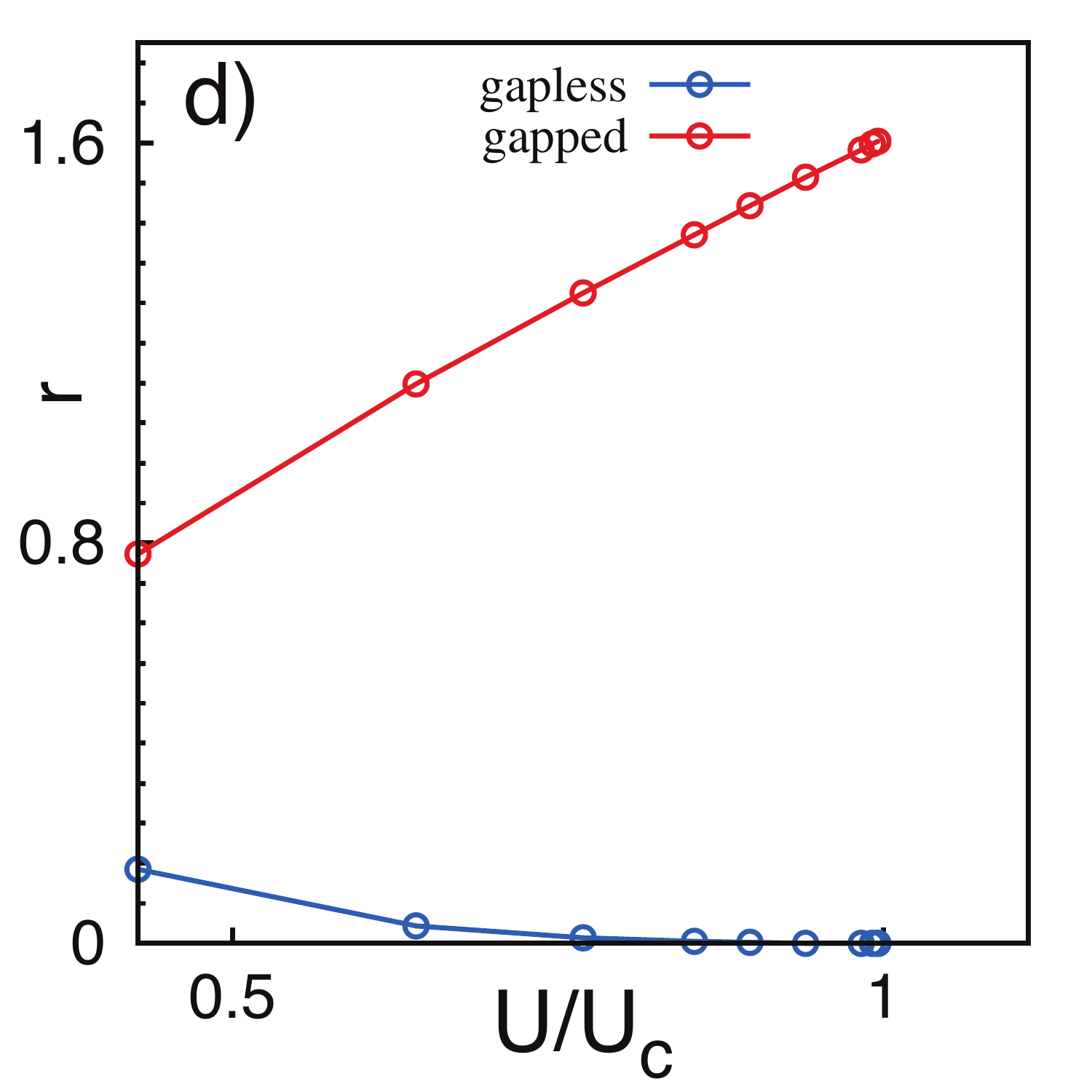}
}
\caption{Key features of the $T=0$ spectrum.
(a)~Superfluid velocity
(b)~difference in energy of the gapless and gapped positive mode
at ${\bf k} = (0,0)~and~(\pi,\pi)$, (c)~dependence of the
residue on interaction strength at ${\bf k} =(0,0)$ and (d)~at 
${\bf k} = (\pi,\pi)$  }
\end{figure}
\begin{figure}[t]
\centerline
{
~~
\includegraphics[width=8.9cm,height=4.2cm]{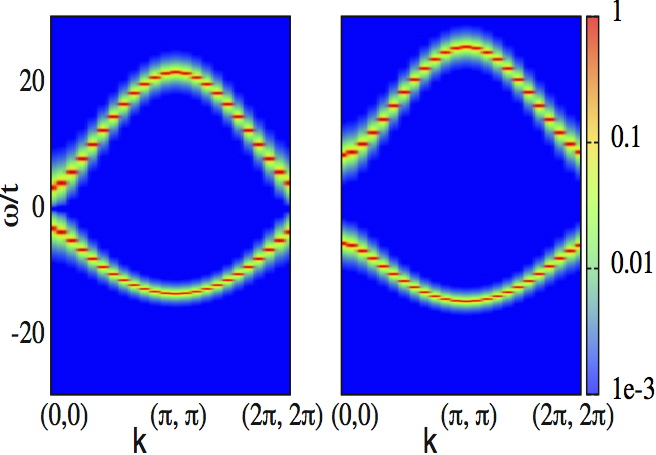}
}
\vspace{.4cm}
\centerline{
~~
\includegraphics[width=3.9cm,height=3.6cm]{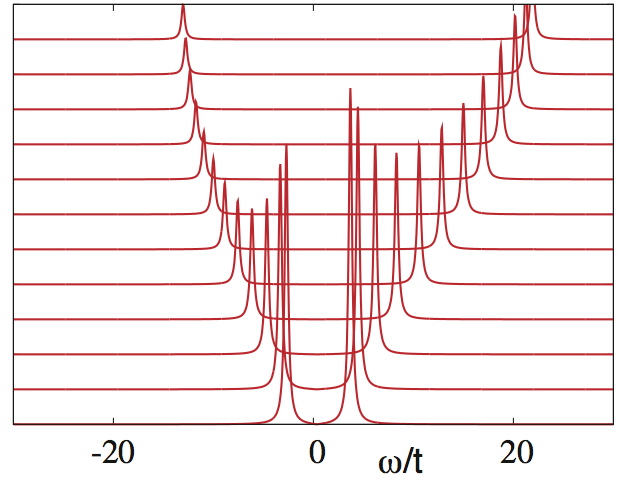}
\includegraphics[width=3.9cm,height=3.6cm]{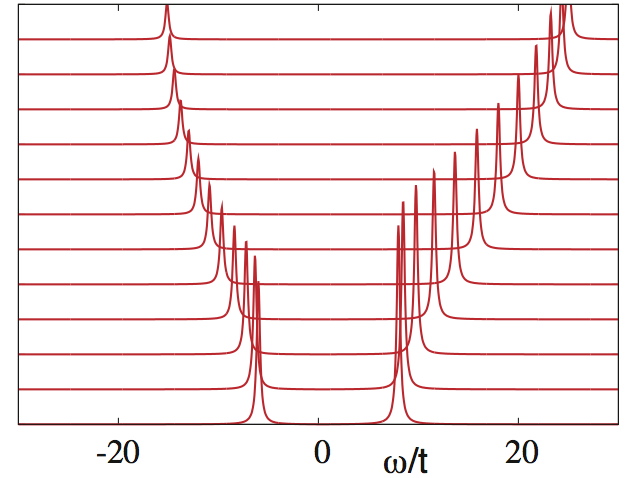}  }
\vspace{.4cm}
\centerline{
\includegraphics[width=4.0cm,height=3.6cm]{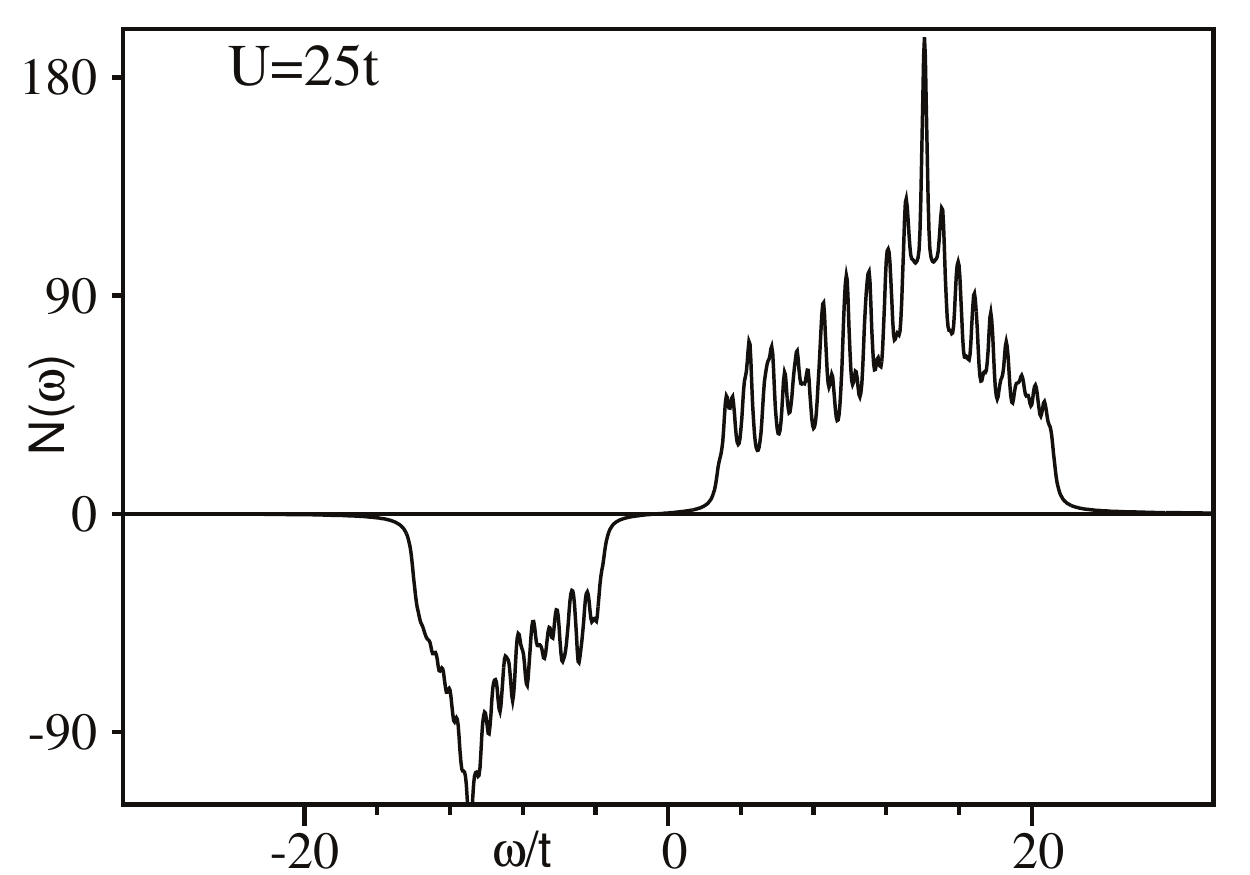}
\includegraphics[width=4.0cm,height=3.6cm]{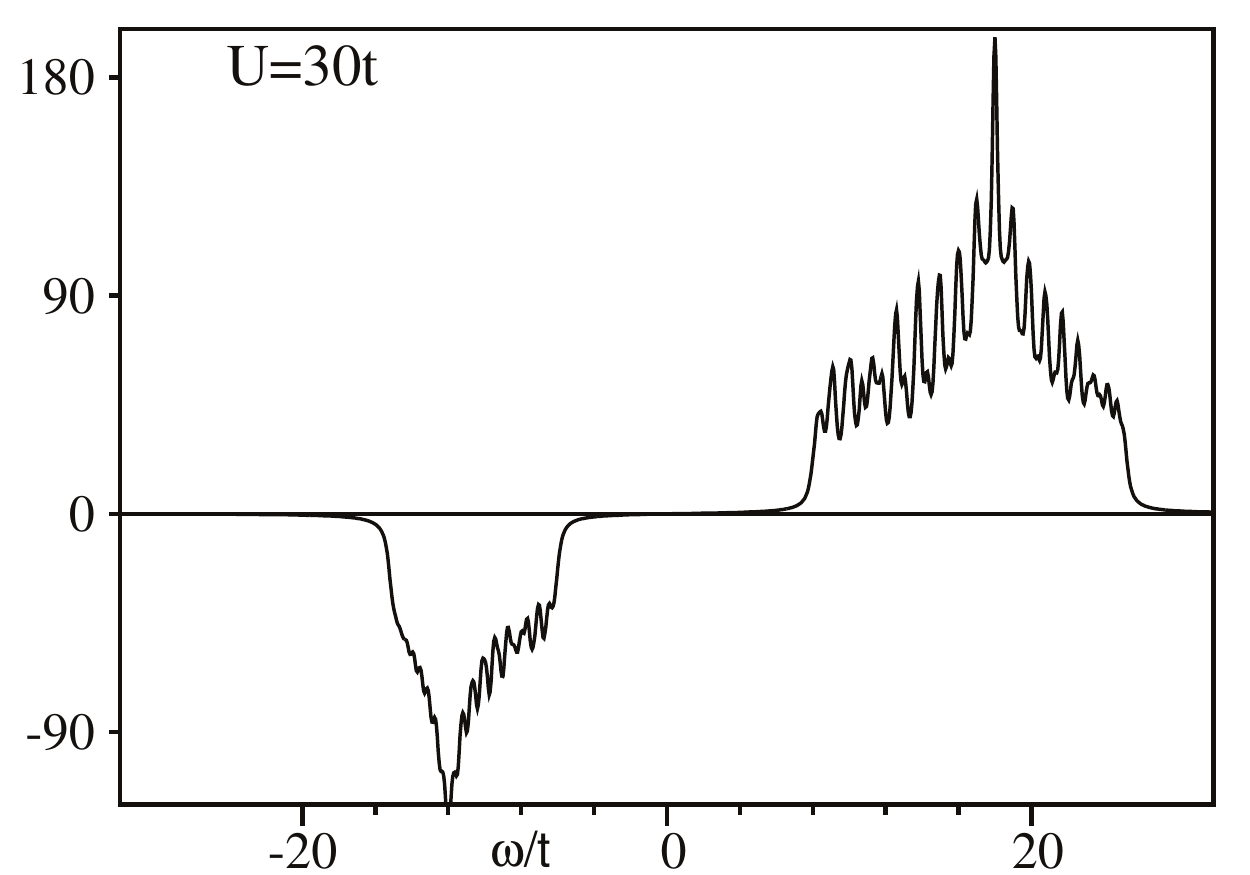} }
\caption{
Spectral function and density of states in the Mott 
ground state. From left to right along each row,
 $U = 25t,~30t$. First row:
$|A({\bf k}, \omega)|$ second row: the lineshape of 
$\vert A({\bf k}, \omega) \vert$,
and  third row: density of states $N(\omega)$.
}
\end{figure}

Looking at the spectral map, top row in Fig.2,
already at $U=2t$ (left panel)
one sees three more bands in additional to
the traditional gapless Bogoliubov band. These include a
negative energy gapless mode, with linear dispersion, and
gapped
amplitude bands of positive and negative energy.
The negative energy gapped band has very small weight and
is visible only near ${\bf k} = (\pi,\pi)$.
Increasing interaction, middle panel, $U=10t$,
makes the weight in the negative energy gapless mode
visibly larger, signifying a broader $n_{\bf k}$
distribution, as well as more prominent amplitude modes.
At $U =22t \sim U_c$, right panel, the gap in the
amplitude modes has almost vanished.
This signifies that in  ``amplitude-phase'' space
the potential energy function is essentially flat and
low momentum fluctuations in either sector, or both,
are of very low energy.

The middle row in Fig.2 shows the lines associated with $\vert A({\bf k}, \omega) \vert$,
which are essentially resolution limited at $T=0$. The momenta values in these
panels increase from $(0,0)$ for the line in the front to $(\pi,\pi)$ for the
line at the back. The patterns reveal the relatively large `mass gap' of the
amplitude mode at $U=2t$ and its progressive reduction as $U$ increases to
$22t$ (where it is still visibly finite). It also shows that at low $U$
and low ${\bf k}$ the only relevant mode is continuum like, since weak
interaction and low ${\bf k}$ are equivalent to the Nozieres-Leggett theory.
However, increasing momentum even at $U=2t$ leads to a sizeable residue
in the amplitude mode, while increasing interaction takes one farther away
from the weak coupling continuum picture.

\begin{figure*}[t]
\centerline{
\includegraphics[width=13.3cm,height=4.0cm]{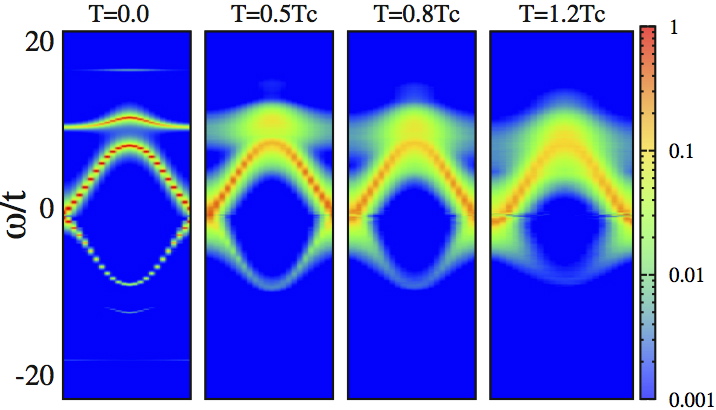}
}
\centerline{
\includegraphics[width=13.3cm,height=4.0cm]{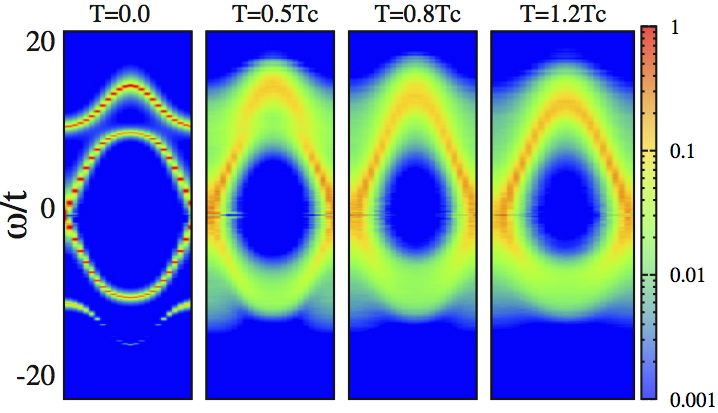}
}
\centerline{
\includegraphics[width=13.3cm,height=4.0cm]{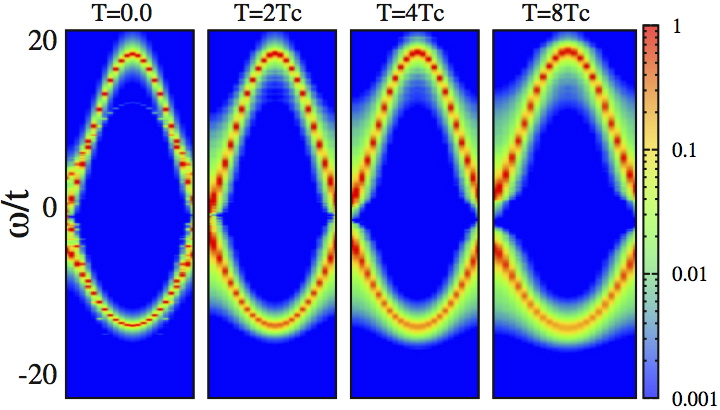}
}
\caption{
$|A({\bf k}, \omega)|$ with increasing temperature for
$U=2t$ (first row), $U=10t$ (second row) and
$U=22t$ (third row). Notice the merger of the Goldstone mode
with the weakly dispersive upper branch as $T$ is increased, and the
loss in weight of the negative frequency branch for $U=2t$ and $10t$.
  For $U=22t$ a  gap opens above $T_c$ and increases with temperature.
}
\end{figure*}

The third row in Fig.2 shows the evolution of
DOS as a function of interaction strength
in the superfluid phase. We find the slope
of DOS around $\omega=0$ increasing as a function
of interaction strength. This
is the mark of increasing superfluid velocity
with $U$.
On the negative frequency axis the DOS
increases with interaction strength. This is related to the
condensate depletion as we argued for the spectral function.
At positive
frequency and small $U$ we see a dip in density of
states and then again a rise. 
This arises from the
separation of the amplitude and phase bands at small $U$.
This dip is not visible at large $U$. 
due to the overlap of the bands.

In Fig.3 we plot some indicators extracted from the
$U$ and ${\bf k}$ dependence of the spectral
functions at more $U$ values than we have shown
in Fig.2. The superfluid velocity,
panel (a), is calculated as the slope of
the dispersion of the positive
energy gapless mode as ${\bf k} \rightarrow 0$.
The slope increases
with $U$. The superfluid velocity almost doubles
from  to $U=2t$ to $U \sim U_c$.
Panel (b) shows the separation between the positive
energy amplitude and phase modes
at ${\bf k}=(0,0)$ (blue) and ${\bf k}=(\pi,\pi)$ (red).
The phase mode is anyway zero energy at ${\bf k}=0$ so the
blue curve basically shows the collapse of the mass gap
of the amplitude mode as $U \rightarrow U_c$.
Panel (c) shows the variation of the small ${\bf k}$ residue
with $U$ for the positive amplitude and phase bands.
The amplitude residue is vanishingly small as $U \rightarrow 0$,
where the phase mode dominates, but overcomes the phase
residue as $U \rightarrow U_c$.
At ${\bf k} =(\pi,\pi)$, however, the amplitude residue is
much larger than the phase residue for the window $0.5U_c < U < U_c$.

\subsection{The Mott ground state}

In Fig.4 the 
first row shows the map of 
$\vert A({\bf k}, \omega) \vert$, the second
row shows the lineshape and the third row the DOS for
a `weak' Mott insulator,  $U=25t$ and a `deep' Mott state, $ U=30t$.

In the Mott phase there are only two modes - 
termed as particle mode at positive frequency
and hole mode at  negative frequency.
At the transition point both the mode are gapless if one 
is at the tip of the Mott lobe otherwise if 
one approaches the transition point from other points on lobes
only one of the mode is gapless.
As one moves into the Mott phase 
both the modes are gapped and
the gap between them at ${\bf k} =0$ increases with $U$.
The weight of the both the modes for fixed interaction strength
decreases as one goes to larger momentum. 
The depletion of weight 
with increasing momentum becomes slower as one goes higher up
in interaction strength. 
If we compare the absolute 
weight of modes at fixed 
momentum for different $U$ values we see a small decrease 
with increase in $U$.

Fig.4 third row shows the plot of density of states 
inside Mott phase at $U=25t$ and $U=30t$.
The gap in DOS grows increases with $U$. The DOS 
at a fixed interaction strength inside Mott phase 
has a free particle like shape with 
van hove like feature on both side of the 
frequency axis. The band width of the DOS 
decreases with $U$.

\section{Thermal behaviour in different interaction regimes}

As one heats the superfluid we see two distinct
types of transition from superfluid to normal
phase.  The normal state
can be gapless or gapped.
We show the gap opening temperature and
dependence of magnitude of gap
on interaction and temperature in
Fig.1.

\begin{figure}[b]
\centerline{
\includegraphics[width=2.7cm,height=3.5cm]{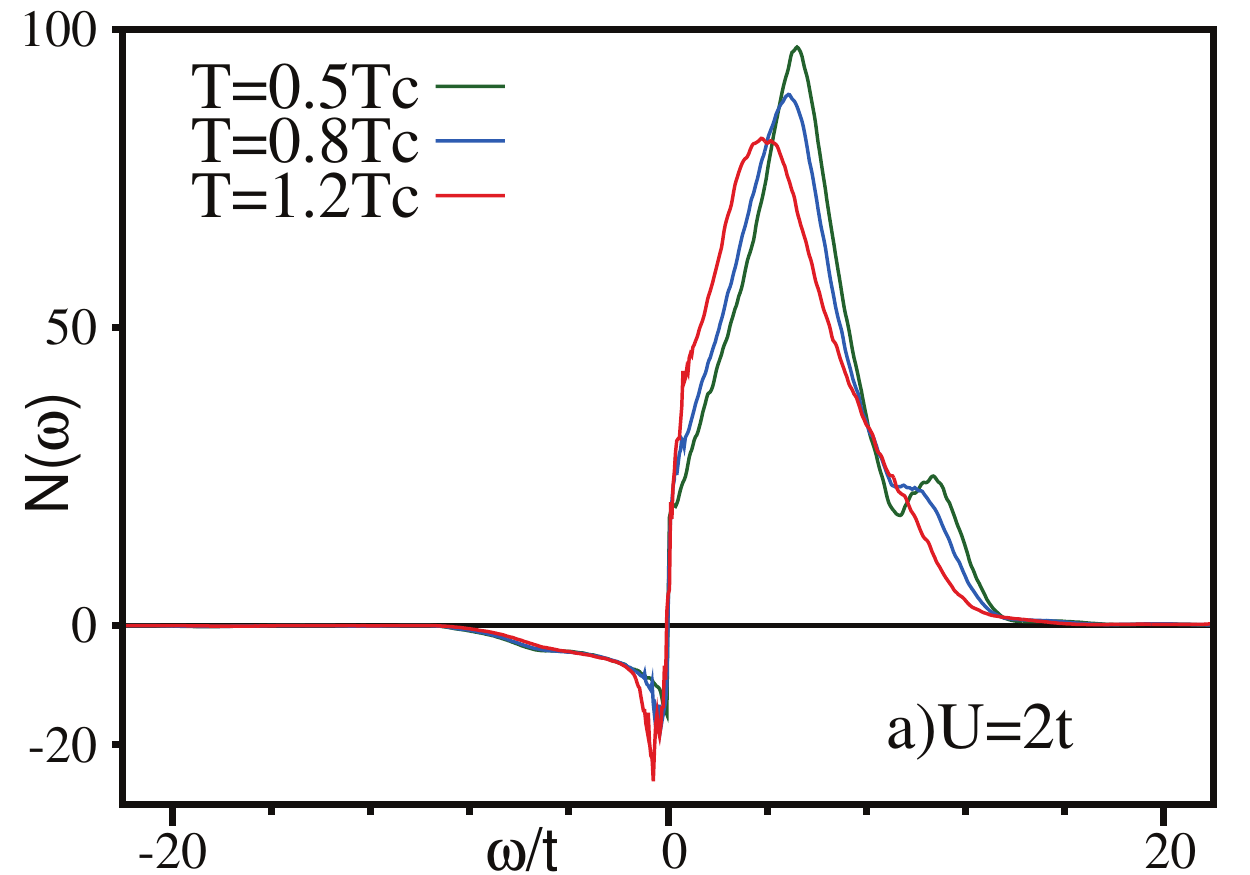}
\includegraphics[width=2.7cm,height=3.5cm]{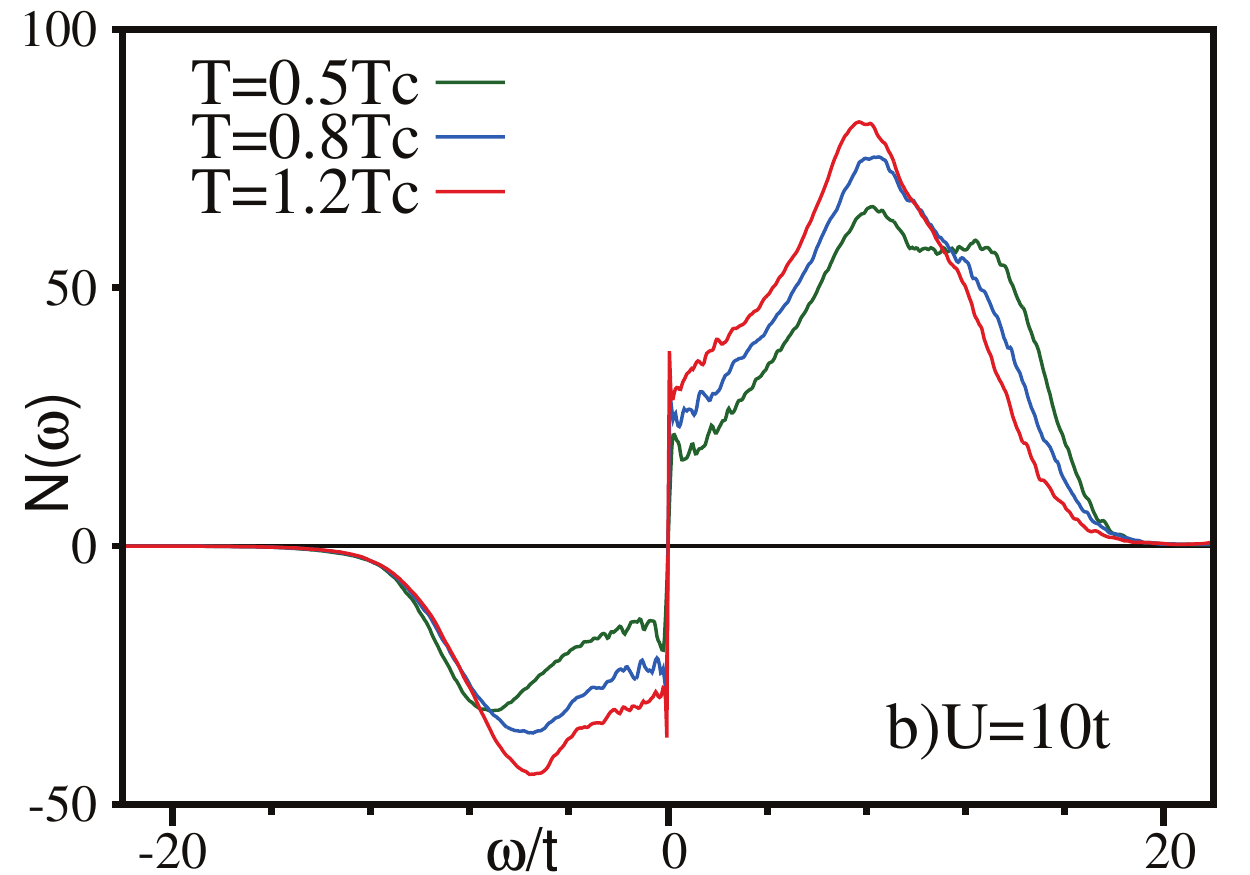}
\includegraphics[width=2.7cm,height=3.5cm]{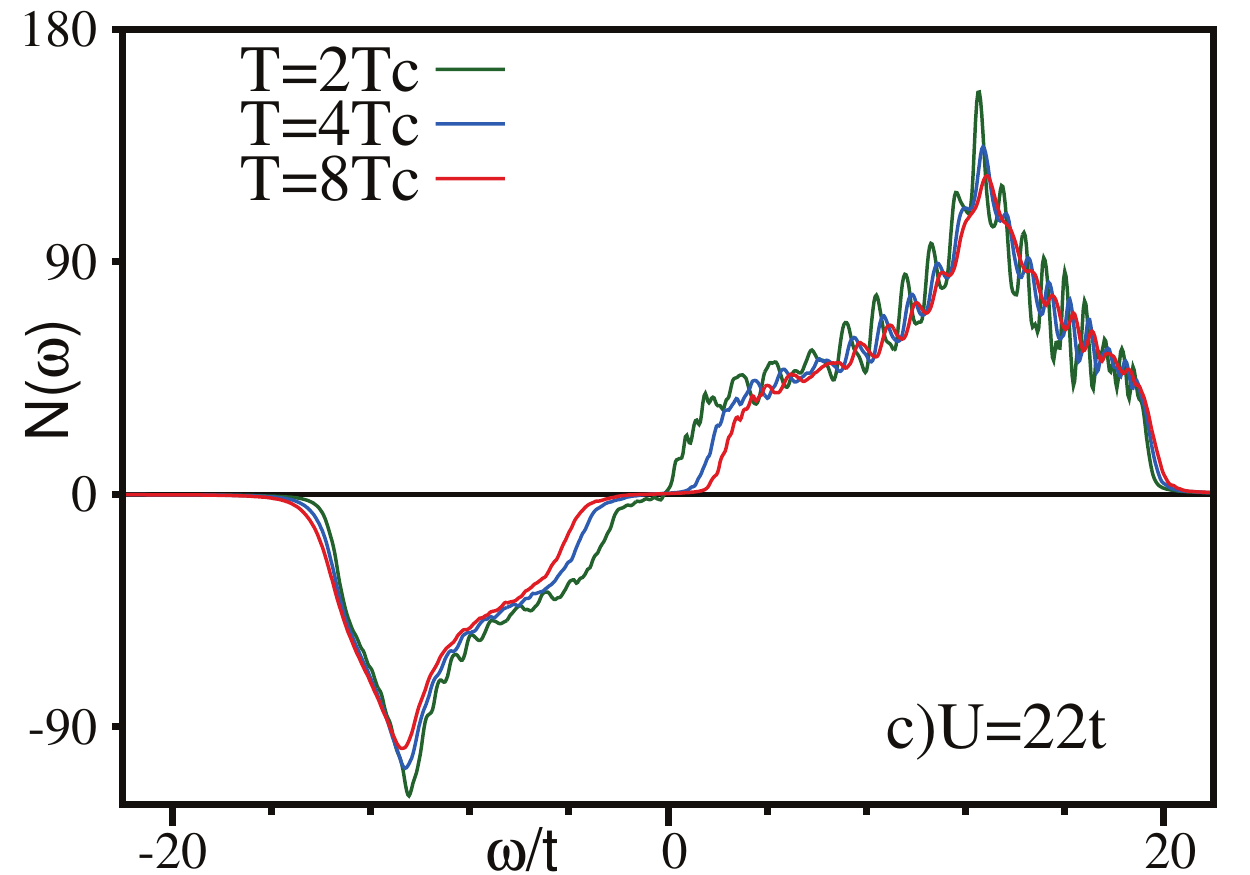}
}
\caption{DOS in the superfluid regime:
$U=2t,~10t,~22t$ for $T/T_c = 0.5,~0.8,~1.2$.Notice 
the opening of gap in DOS for U=22t.   }
\end{figure}
\begin{figure}[t]
\centerline
{
\includegraphics[width=2.8cm,height=3.2cm]{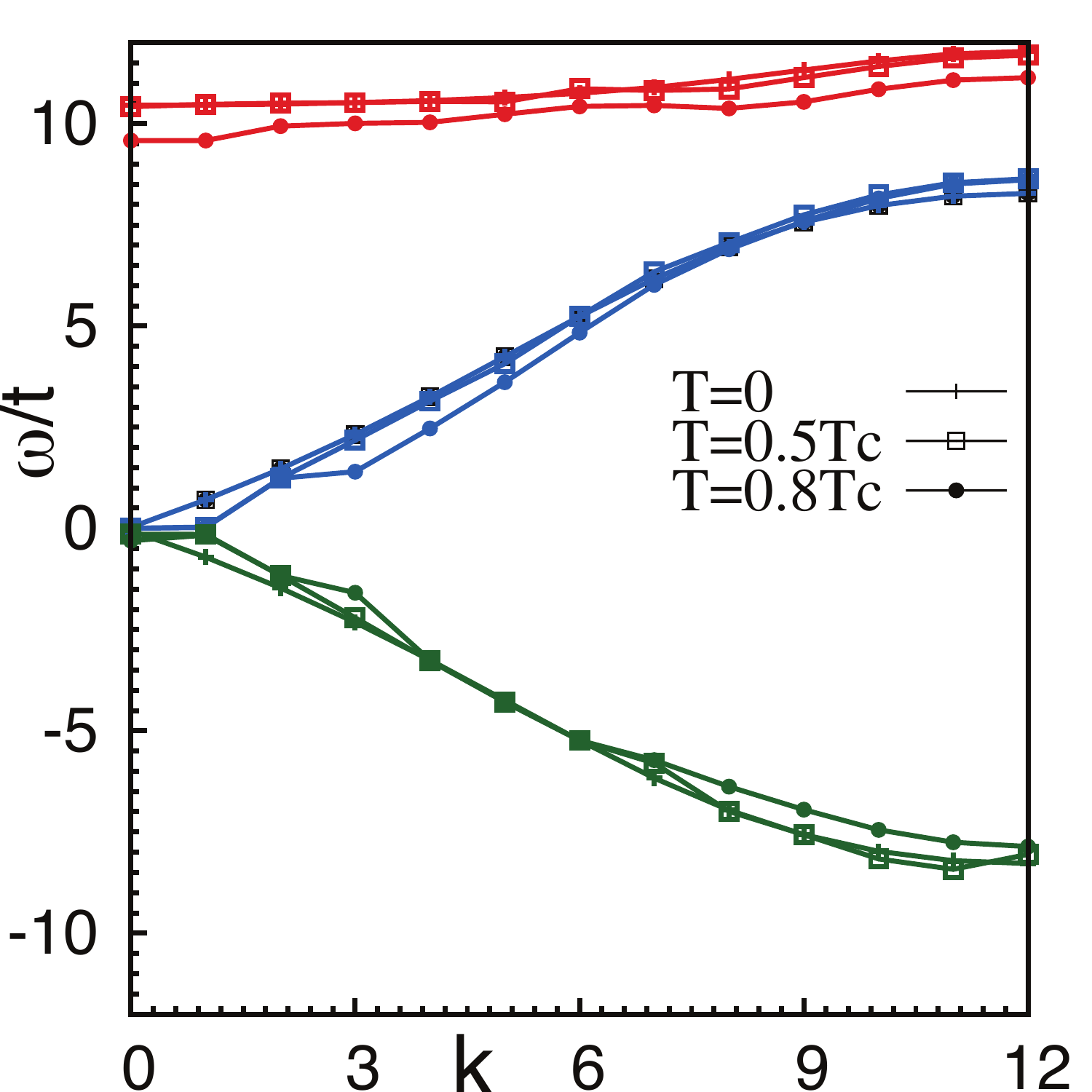}
\hspace{-.2cm}
\includegraphics[width=2.8cm,height=3.2cm]{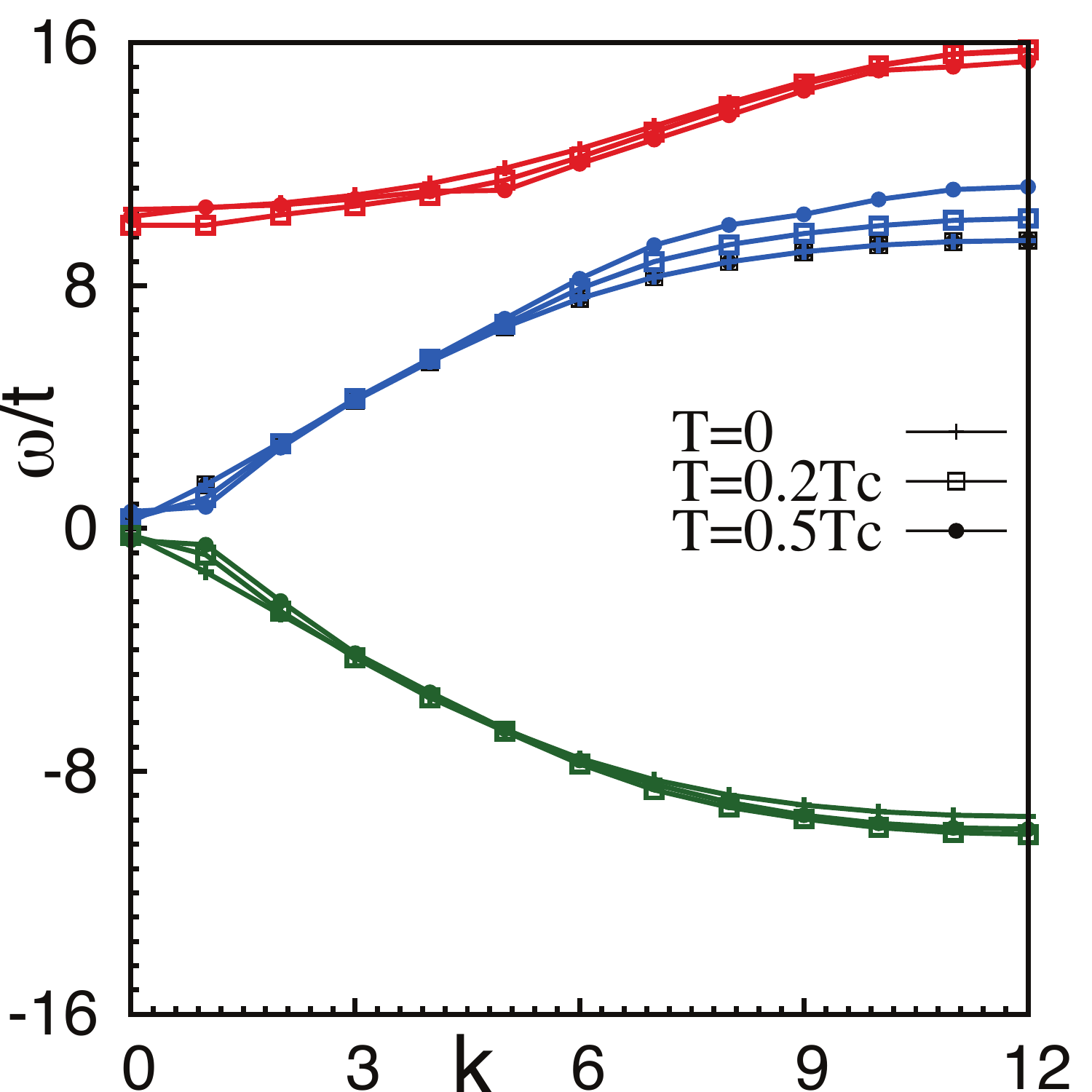}
\hspace{-.2cm}
\includegraphics[width=2.8cm,height=3.2cm]{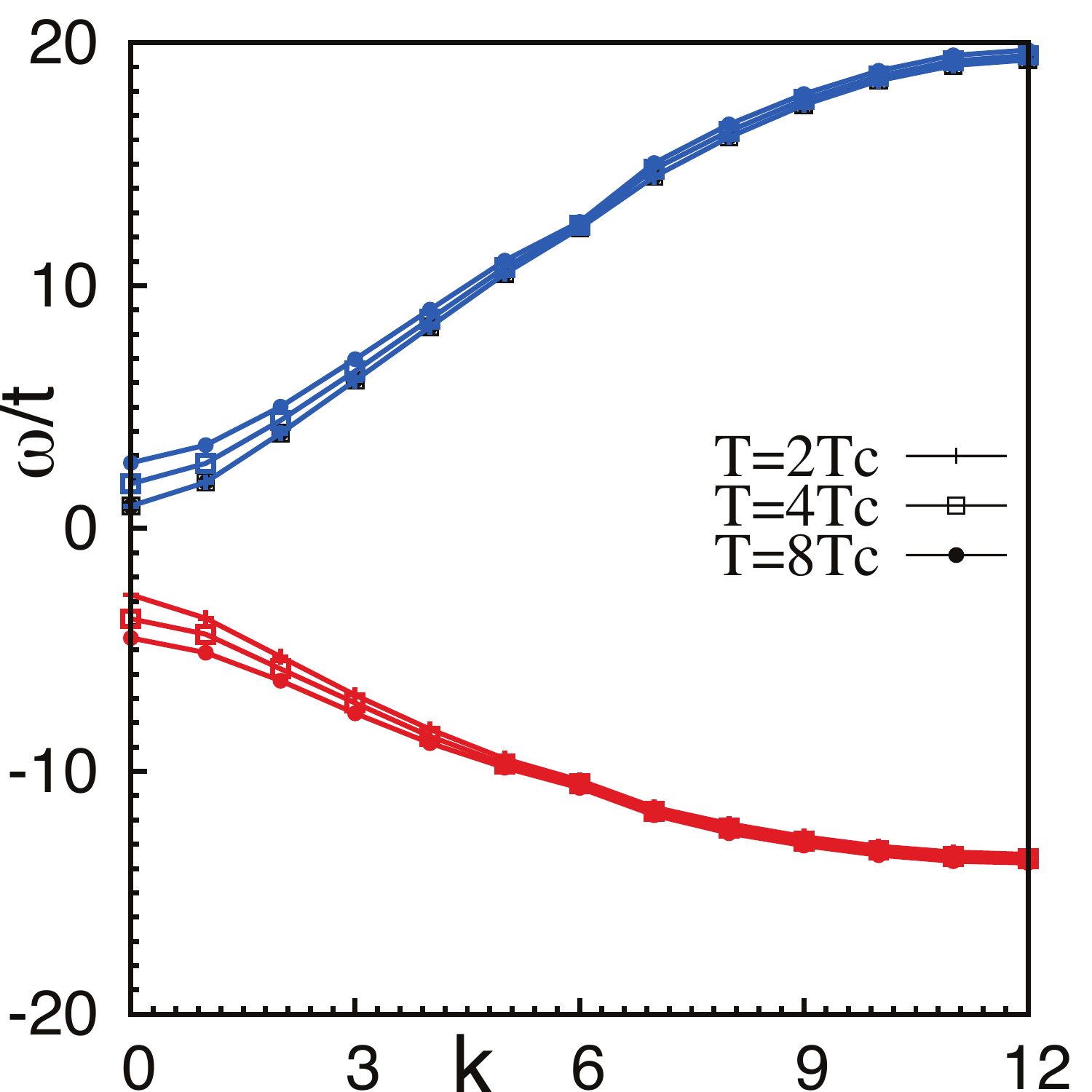} }
\vspace{.3cm}
\centerline
{
\includegraphics[width=2.8cm,height=3.2cm]{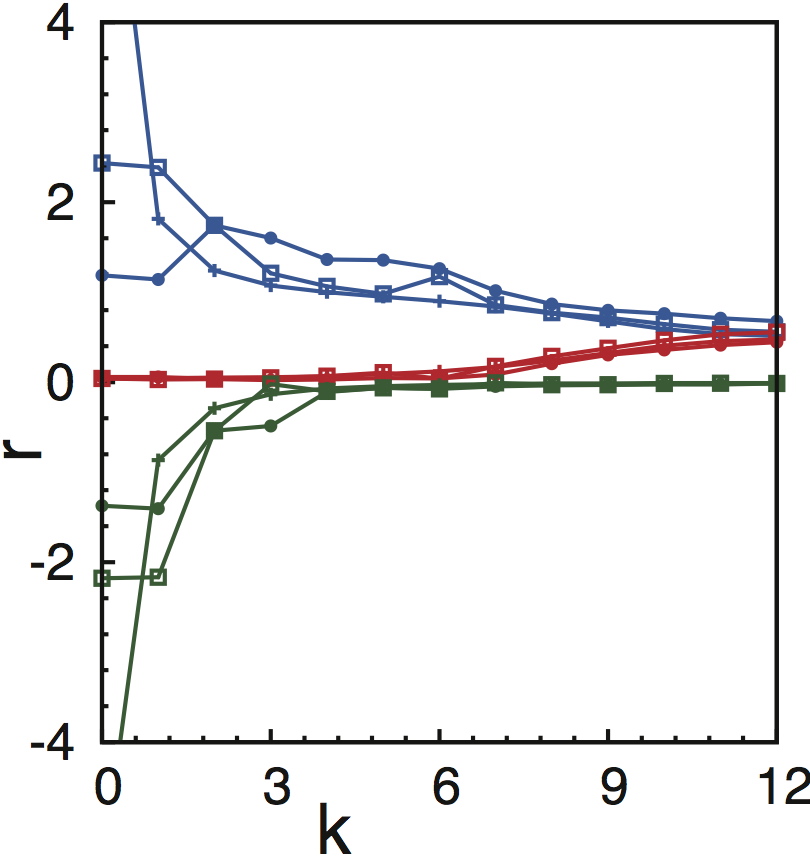}
\includegraphics[width=2.8cm,height=3.2cm]{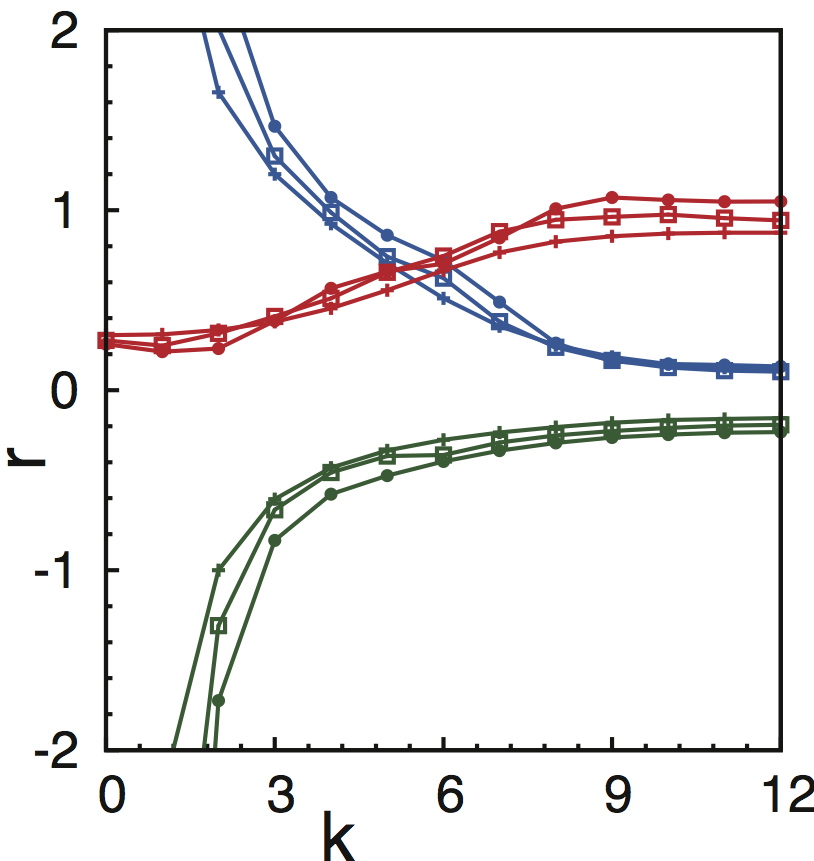}
\includegraphics[width=2.8cm,height=3.2cm]{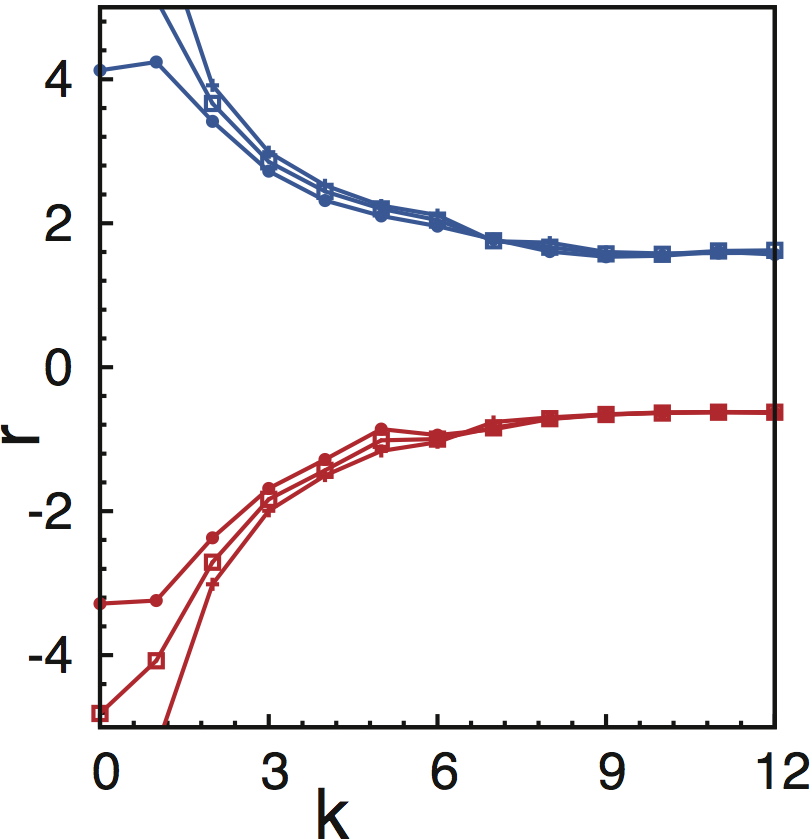} }
\centerline
{
~
\includegraphics[width=2.7cm,height=3.2cm,angle=0]{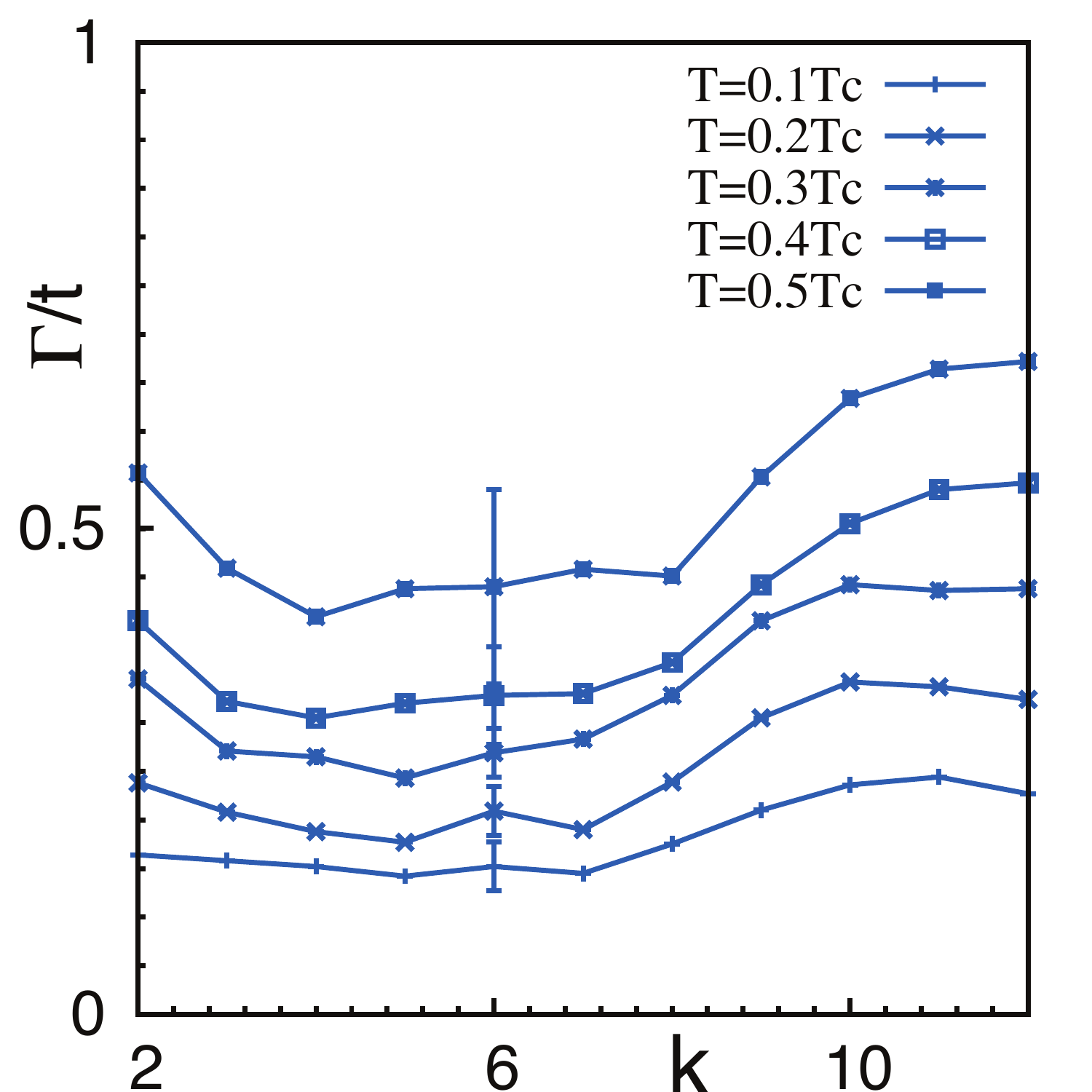}
\includegraphics[width=2.7cm,height=3.2cm,angle=0]{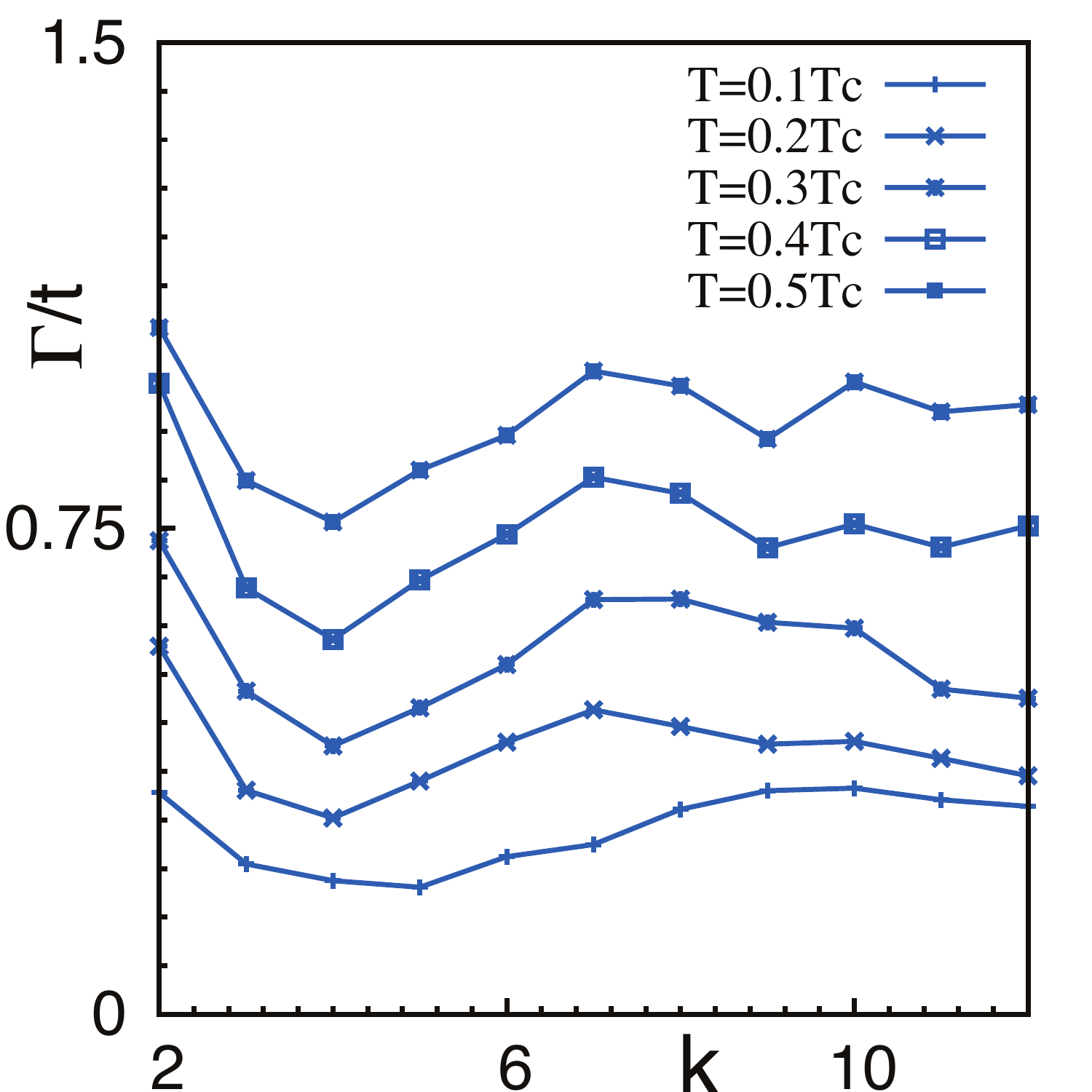}
\includegraphics[width=2.7cm,height=3.2cm,angle=0]{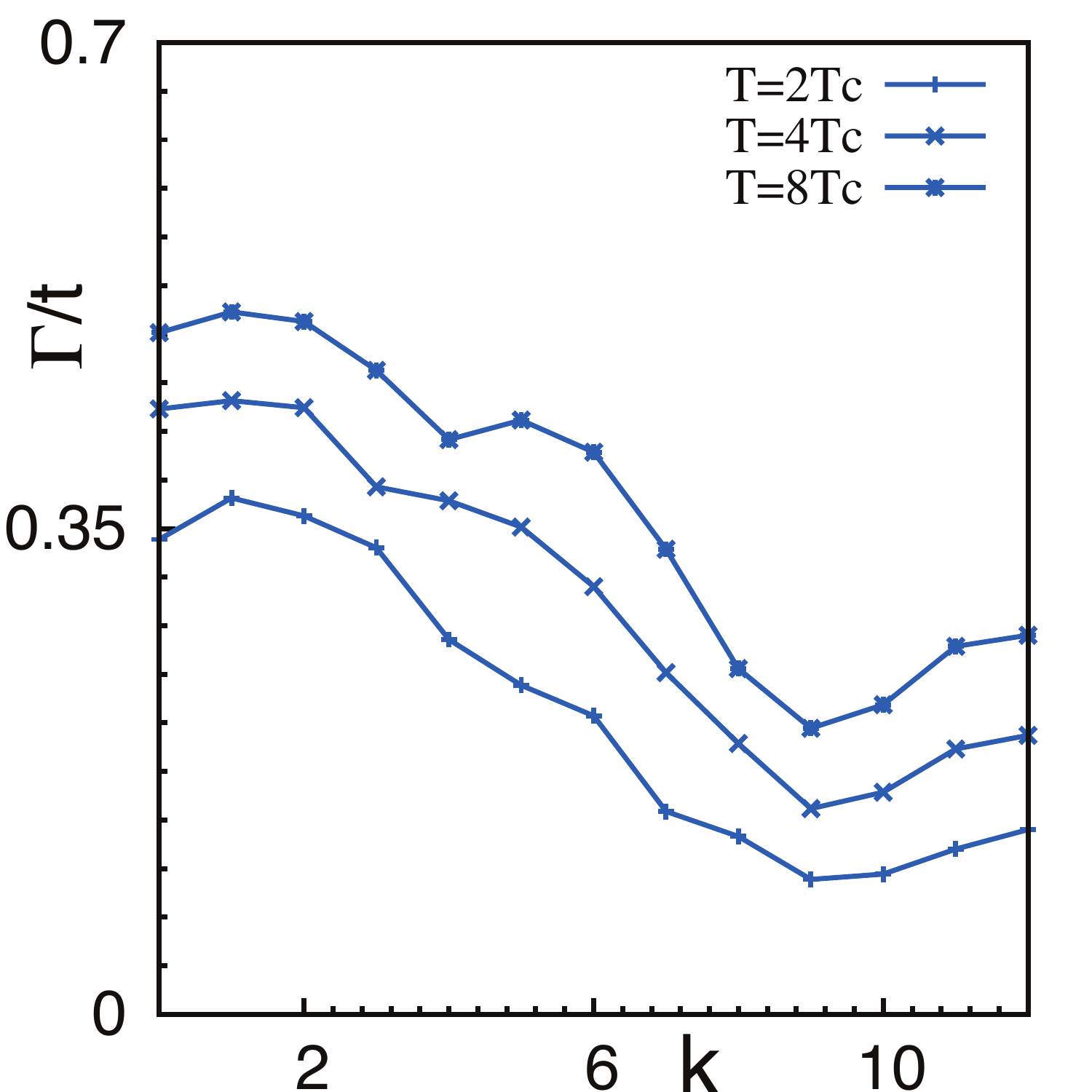}
}
\vspace{.3cm}
\centerline
{
~
\includegraphics[width=2.7cm,height=3.2cm,angle=0]{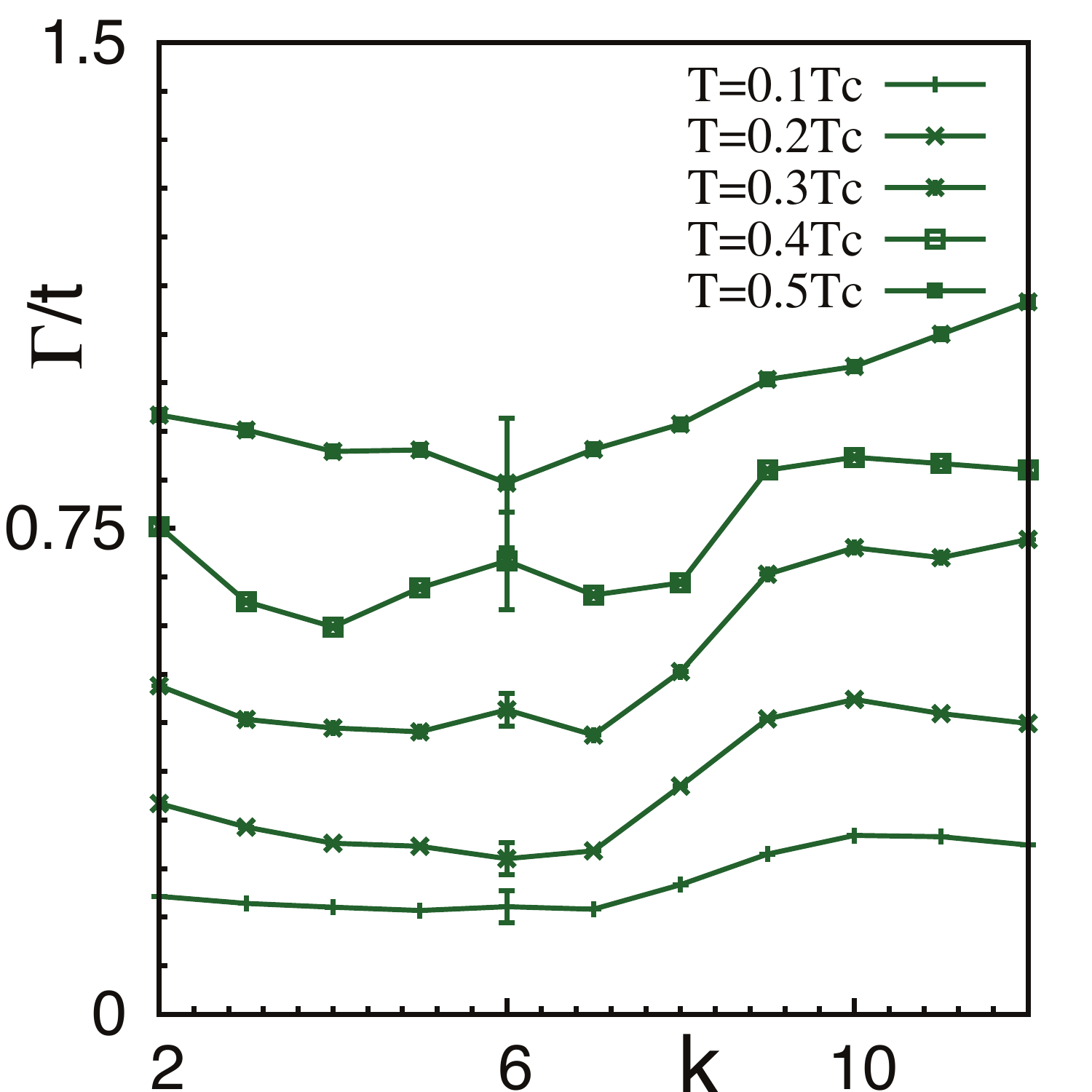}
\includegraphics[width=2.7cm,height=3.2cm,angle=0]{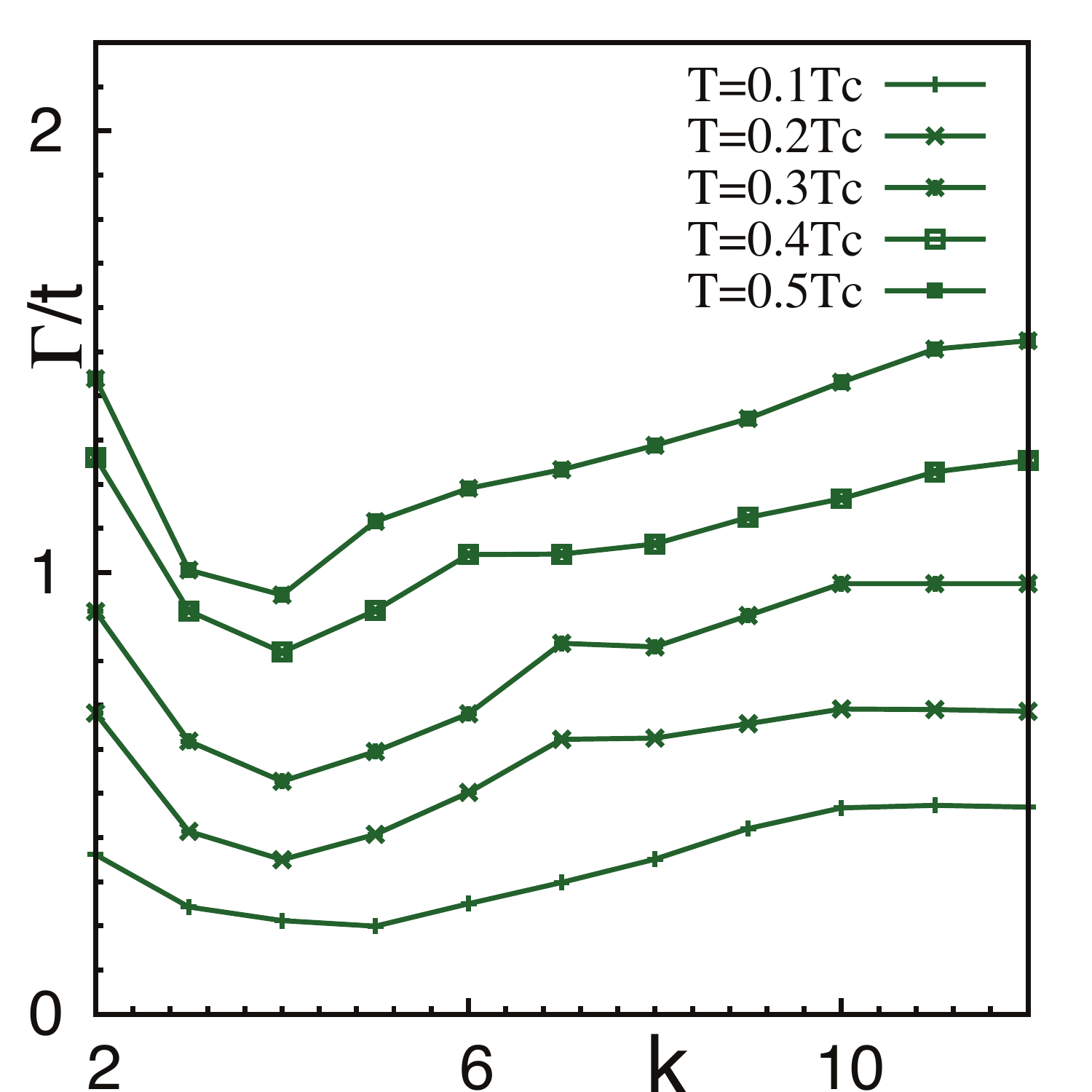}
\includegraphics[width=2.7cm,height=3.2cm,angle=0]{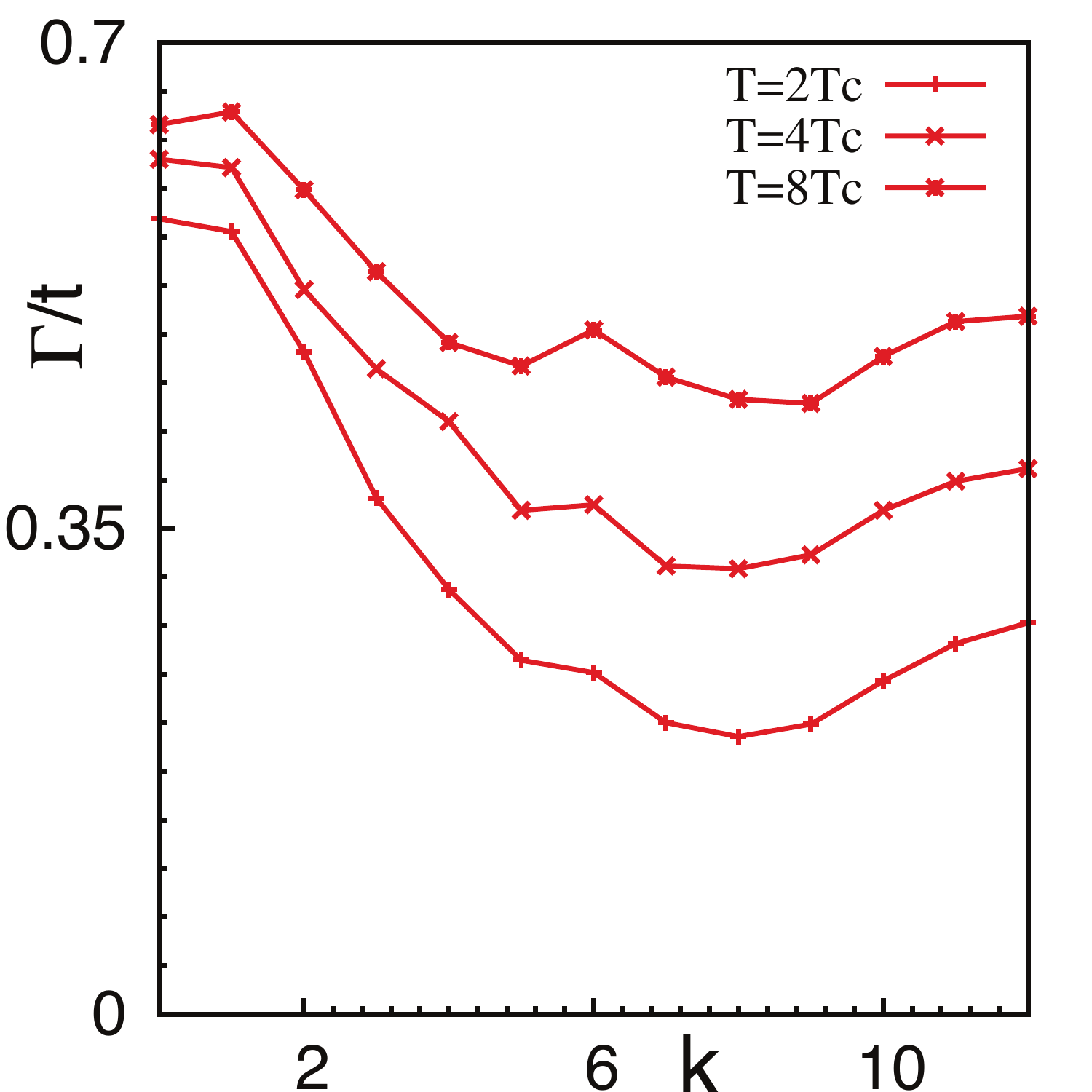}
}
\caption{  Parametrisation of the finite $T$
spectral data in the superfluid regime.First row:
mean dispersion, second row: residue,third row:broadening 
for the gapless band on positive frequency axis,
fourth row :broadening for gapless band on negative frequency axis.
From left to right, U goes from
 $2t, 10t, 22t.$  }
\end{figure}

\subsection{The superfluid at finite temperature}

Fig.5 shows the plot of $\vert A({\bf k}, \omega) \vert$
for $U/t = 2, 10, 22$ and varying $T$.
First we discuss the results for $U=2t$ and $10t$.
At zero temperature the gapless
and gapped modes are sharply peaked
function of $\omega$
for each wave vector. The
increase in  temperature
leads to reduction in weight and broadening of
modes. There is a huge reduction in weight
for small momenta, while the residue at 
large momenta remain largely $T$ independent.
We find that for $U=10t$ beyond $T=0.5T_c$
there is not much reduction in weight.
In general the reduction is more prominent on
the negative frequency axis.

At low temperature only 
small momenta have large non-zero weight
close to $\omega\sim0$.
This changes with temperature rise as other wave
vectors now have large non zero weight at
frequency close to zero and with a long tail at large
$\omega$. This gets more pronounced as
one goes up in temperature and more prominent for the 
small $U$ superfluid.
For $U=2t$ it is possible to distinguish between gapless
and gapped band for temperatures less than T$_c$
but for $U=10t$ even above $0.5T_c$ this distinction
is hard to make due to thermal broadening. This
feature gets worse as one goes higher up in interaction
strength since at zero temperature gapless and gapped
bands are close to one another. Above a temperature
it is difficult to discuss in terms of multiple band
but the single band picture emerges due
to band overlapping.

\begin{figure*}[t]
\centerline
{
\includegraphics[width=12.3cm,height=3.7cm]{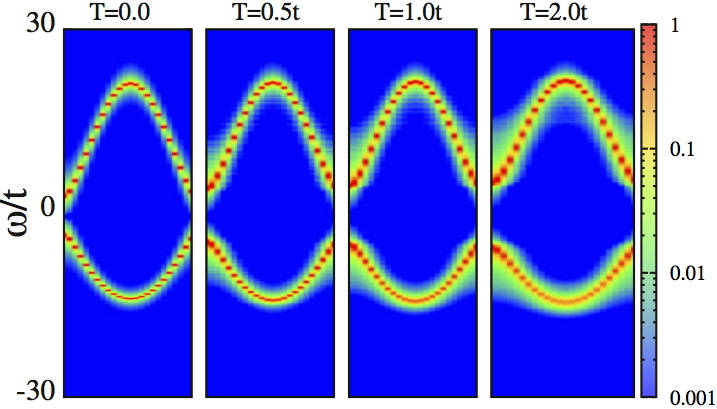}
}
\centerline
{
\includegraphics[width=12.3cm,height=3.7cm]{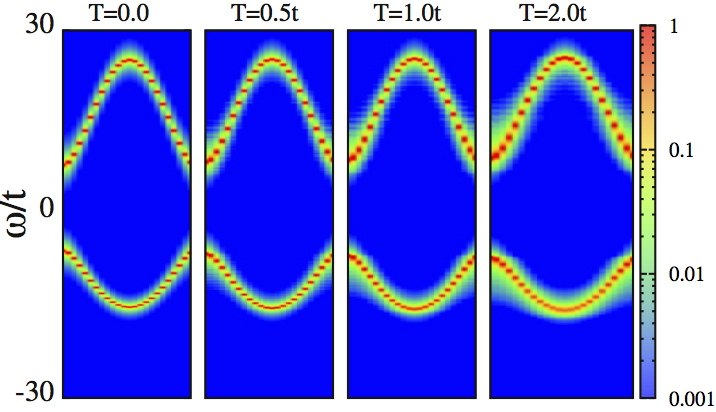}
}
\caption{
Spectra in the finite $T$ Mott state.
$|A({\bf k}, \omega)|$ with increasing temperature as a 
fraction of $t$. First row:$U=25t$, second row: $U=30t$.}
\end{figure*}

Now we discuss superfluid close to the transition, $U=22t$.
At zero temperature excitation spectra
consists of gapless and gapped mode.
The temperature rise leads to broadened
spectra with reduction in weight
ultimately to creation of gap in the
excitation spectra which keeps increasing
with temperature.  With rising temperature
the suppression and shifting of peak to
large frequency is seen for small momentum.

\begin{figure}[b]
\centerline{
\includegraphics[width=4.2cm,height=3.5cm]{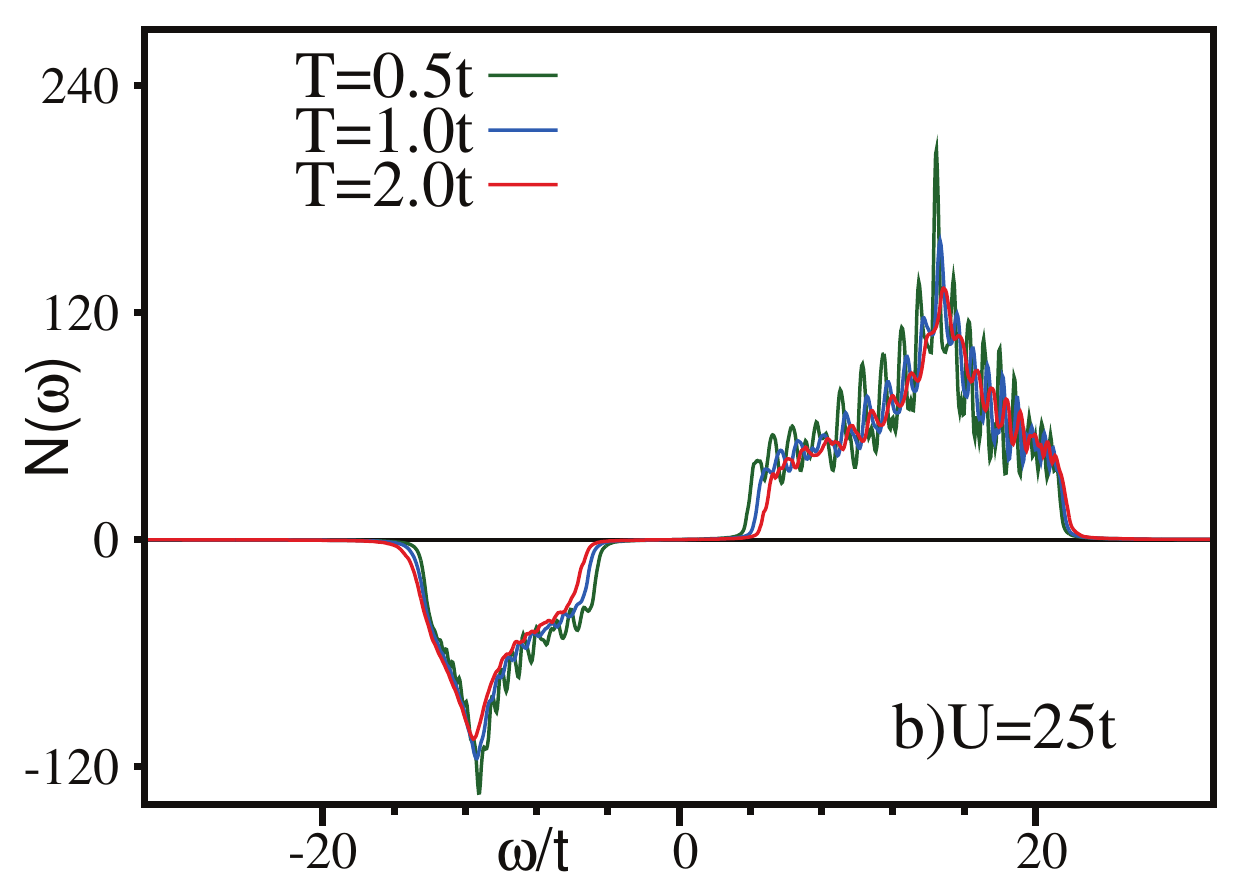}
\includegraphics[width=4.2cm,height=3.5cm]{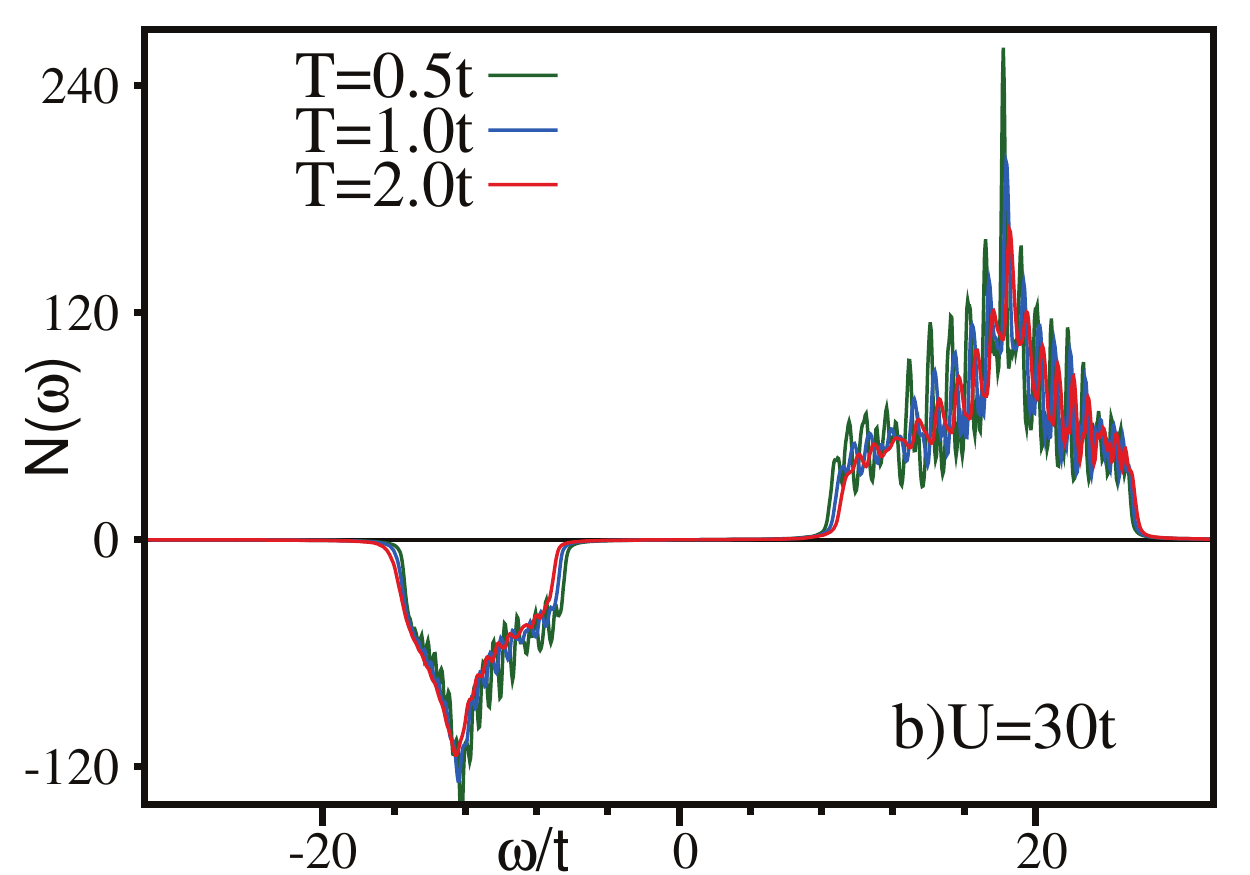}
}
\caption{DOS in the finite $T$ Mott phase.The increase 
in gap and suppression of van-hove like peak with temperature 
is  visible  }
\end{figure}

The 
broadening
is visible and two peak structure at small temperature
is converted to single peak structure
at finite large temperature.  For large
momentum only thermal
effect is suppression of peak with its
position remaining unchanged.The
percentage suppression is large for small
momentum as compared to large momentum.
At large finite temperature
the peak in $A({\bf k}, \omega)$ 
has non monotonic dependence
on momentum.
Below $T=0.5t$ it is difficult to
distinguish between the phonon
and amplitude band but above $0.4t$
only single band is visible so we show
data for temperature greater
or equal to 0.5t.

\begin{figure}[b]
\centerline{
~
\includegraphics[width=3.5cm,height=2.5cm,,angle=0]{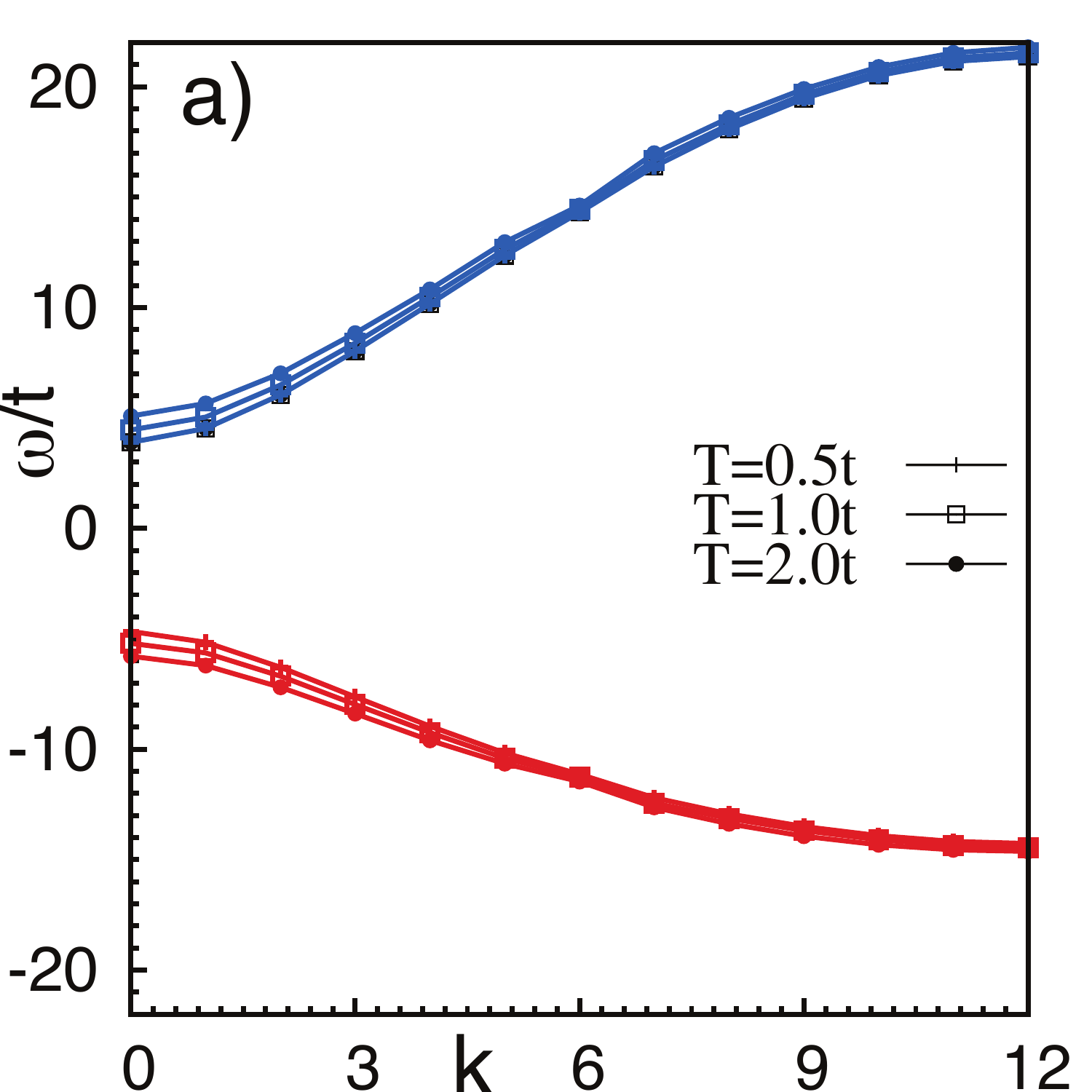}
\hspace{.2cm}
\includegraphics[width=3.5cm,height=2.5cm,angle=0]{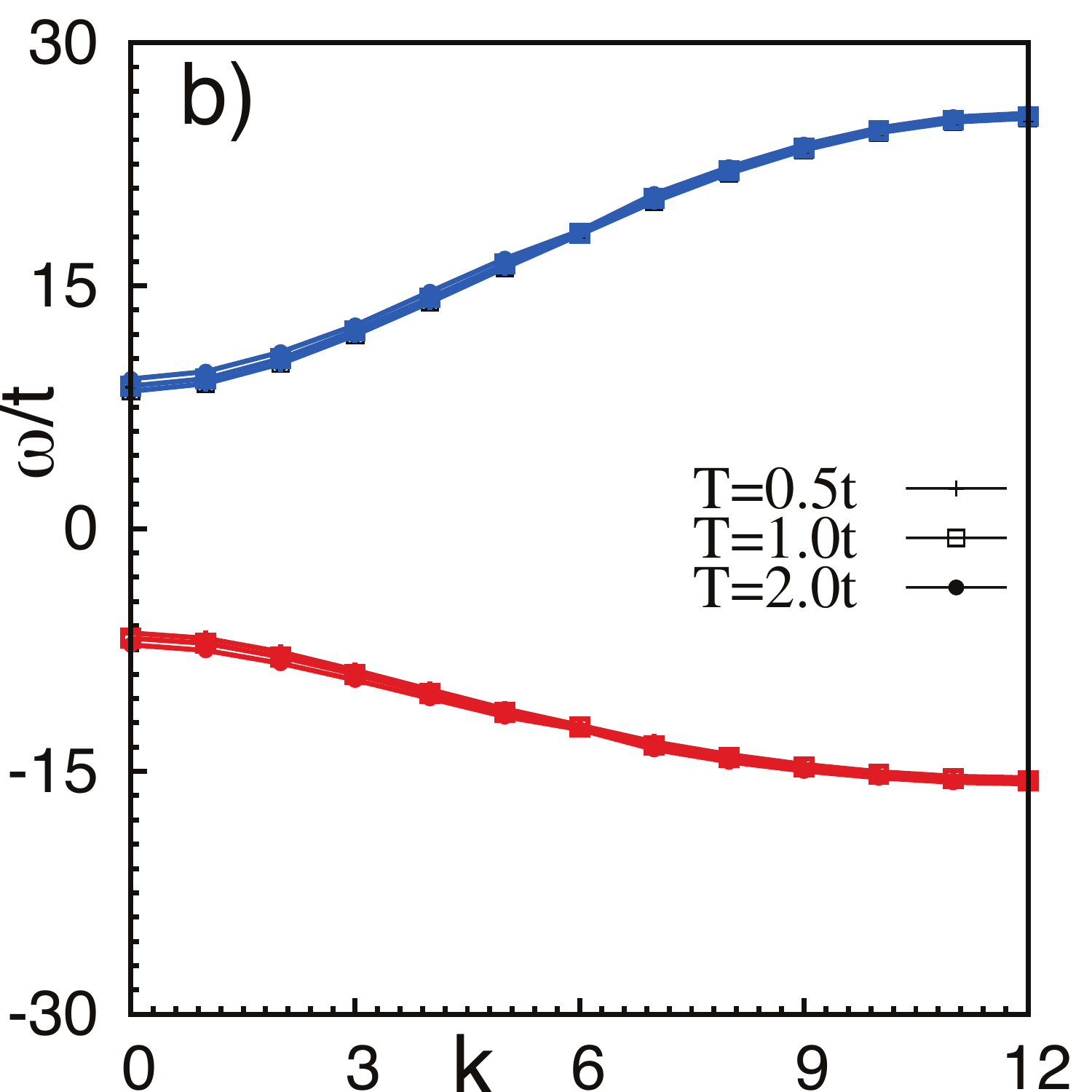}
}
\vspace{.1cm}
\centerline{
\includegraphics[width=3.7cm,height=2.5cm,angle=0]{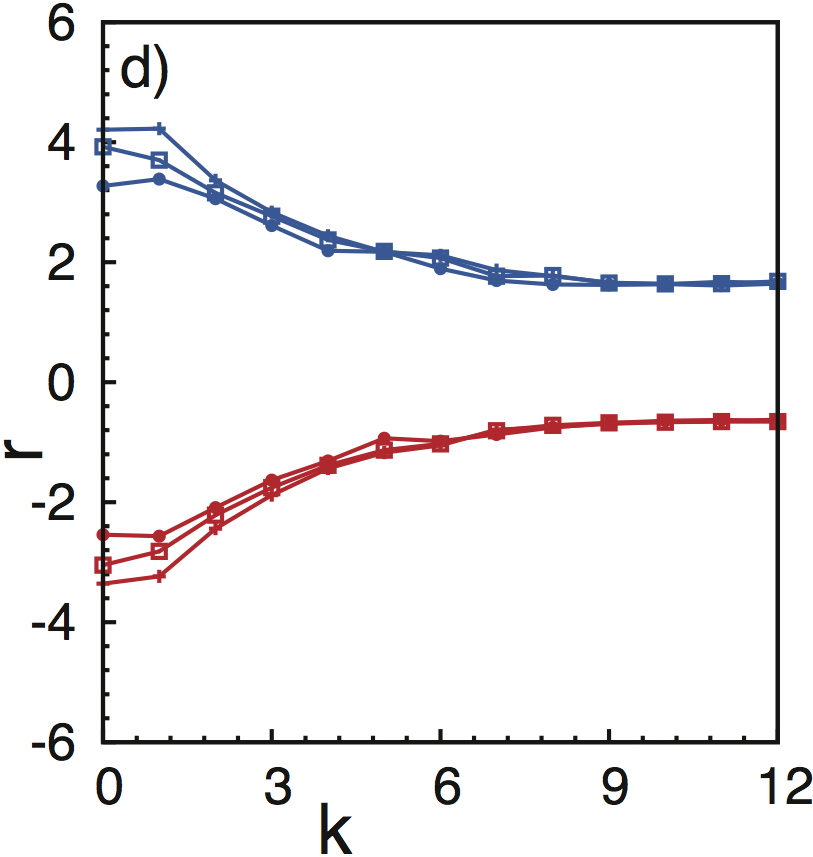}
\includegraphics[width=3.7cm,height=2.5cm,angle=0]{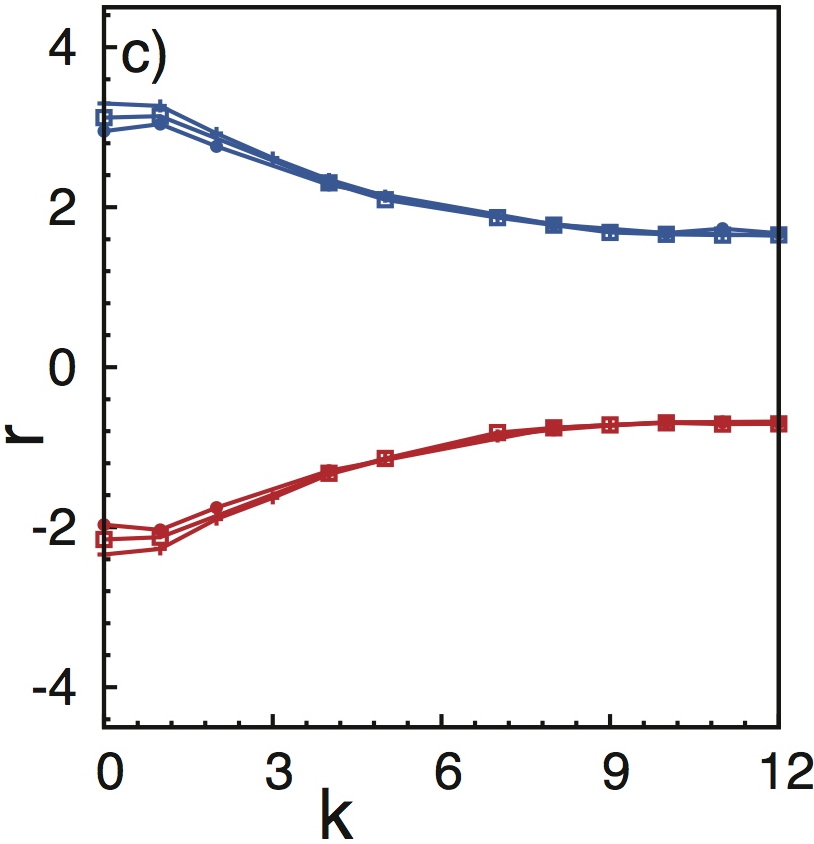}
}
\centerline{
~
\includegraphics[width=3.6cm,height=2.5cm,,angle=0]{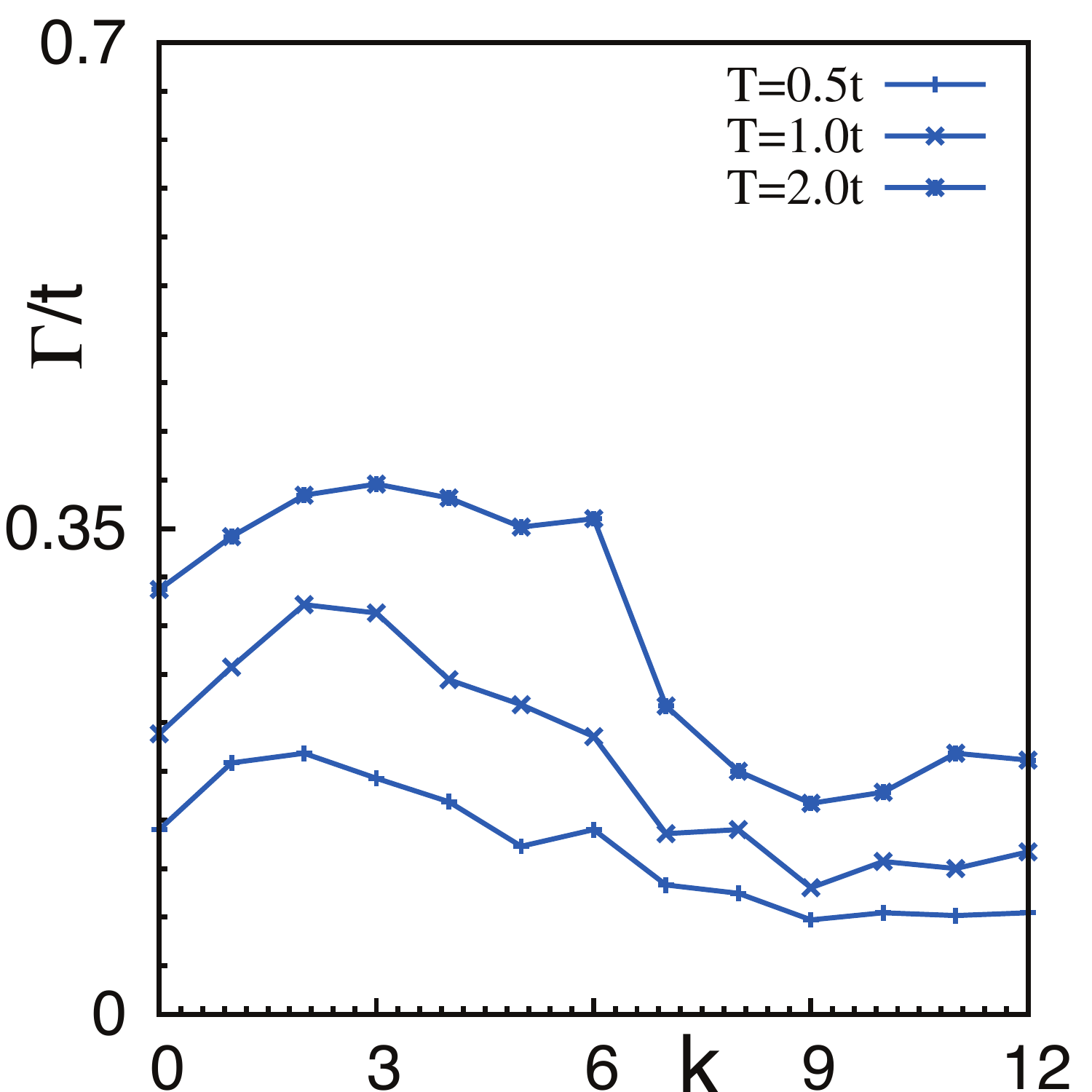}
\includegraphics[width=3.6cm,height=2.5cm,,angle=0]{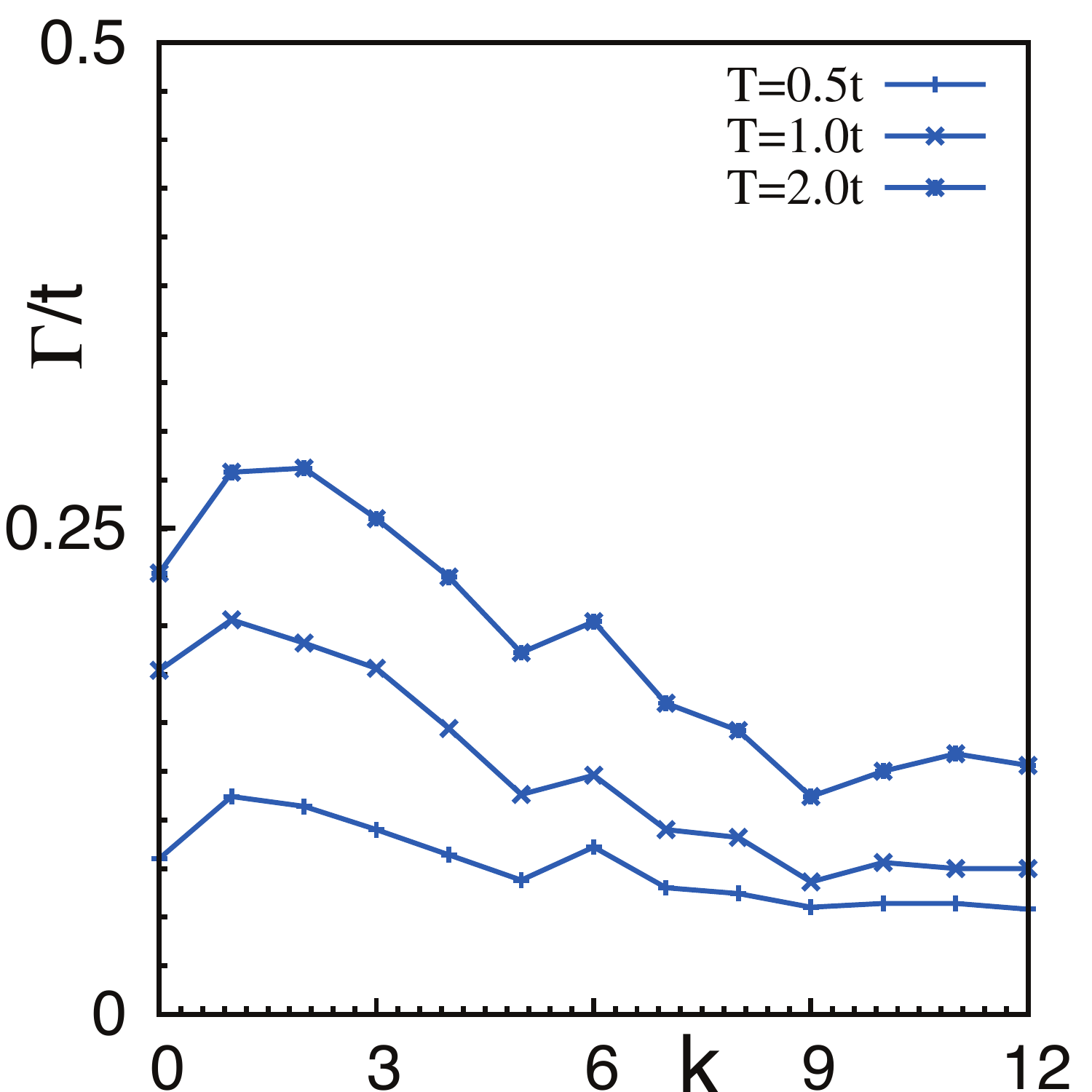}
}
\centerline{
~\includegraphics[width=3.6cm,height=2.5cm,,angle=0]{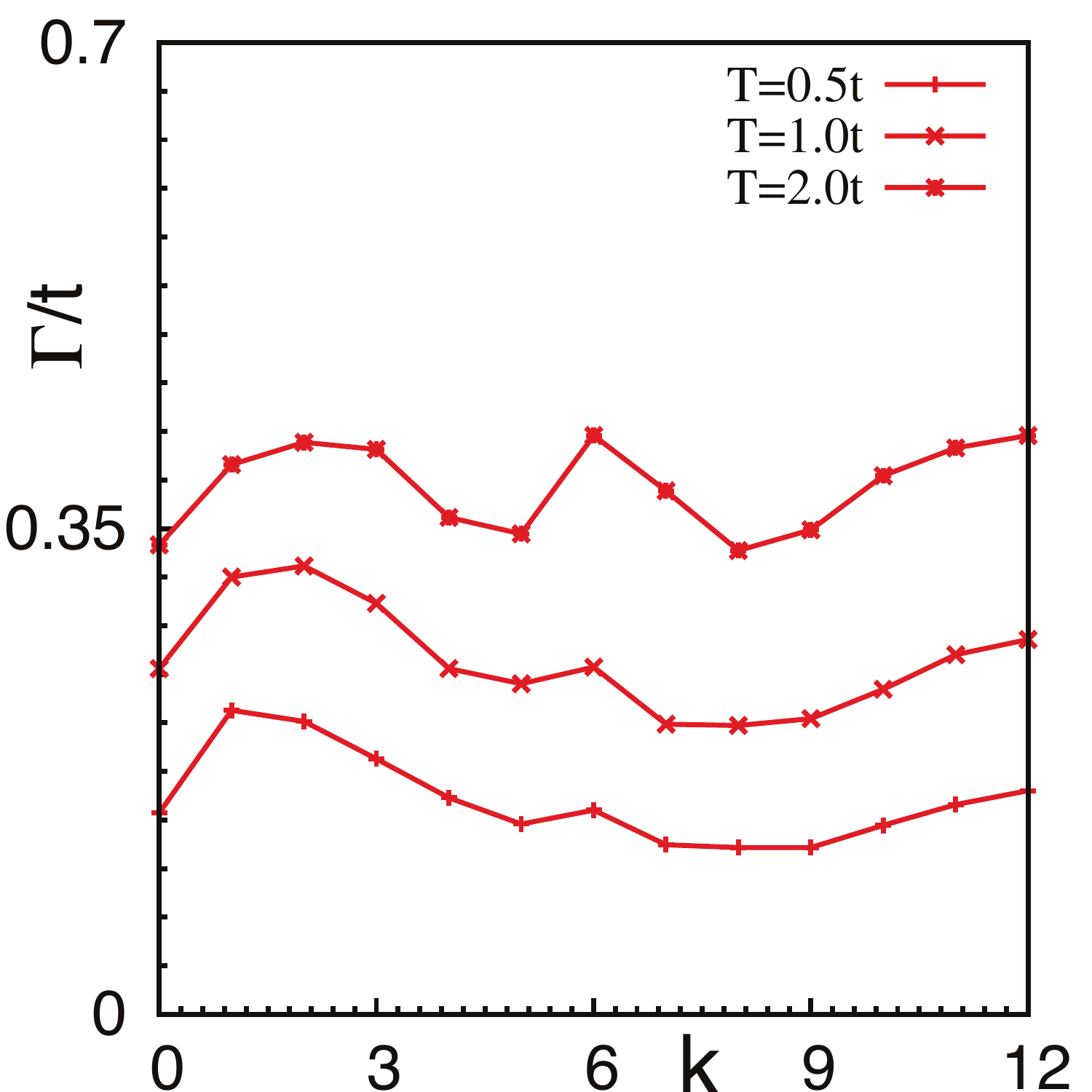}
\includegraphics[width=3.6cm,height=2.5cm,,angle=0]{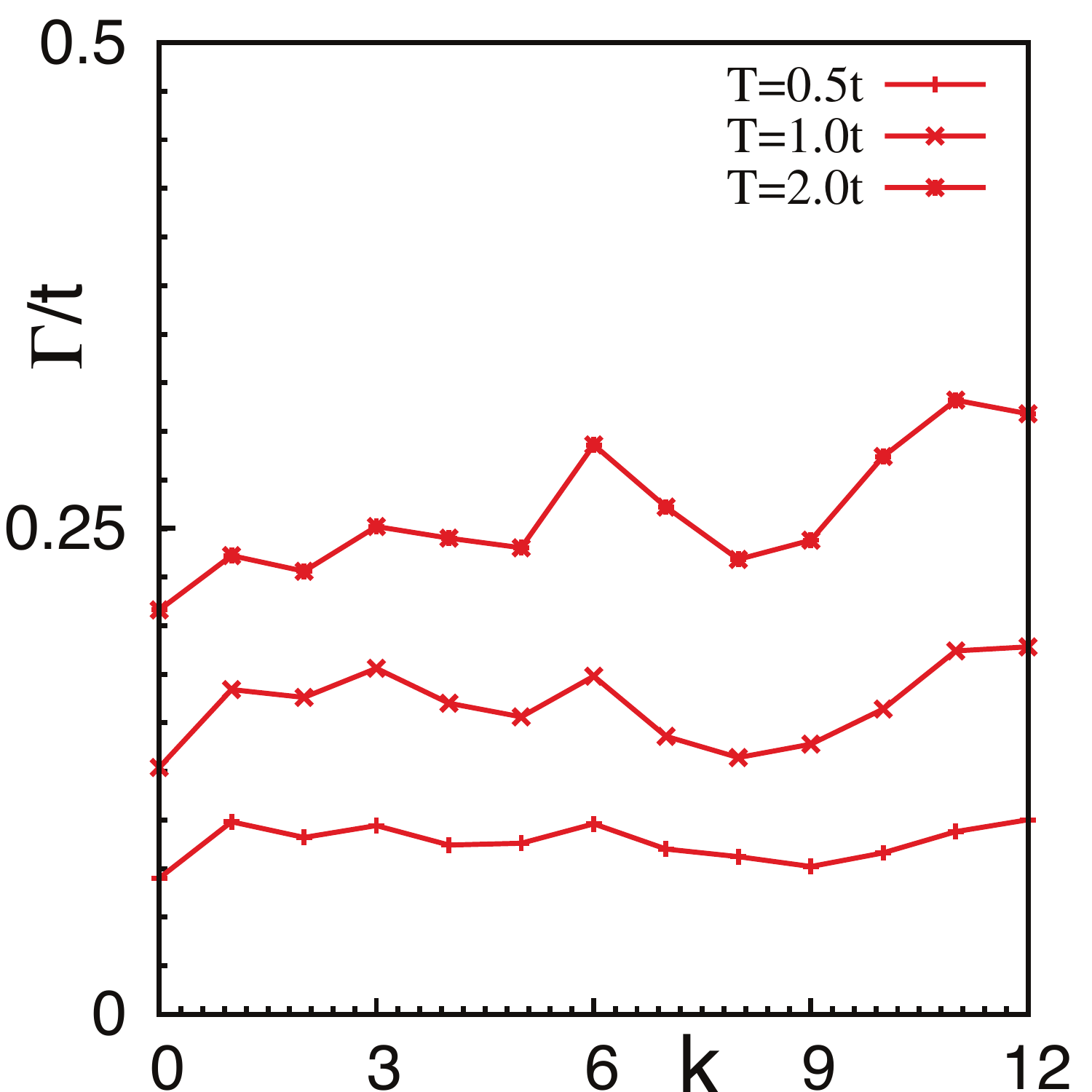}
}
\caption{Parametrisation of the spectral data.
First row:
mean dispersion, second row: residue,third row:broadening 
for the particle band,
fourth row :broadening for hole band.
Left: $U=25t$, right: $U=30t$.
}
\end{figure}

We track the average dispersion and broadening of
the bands by fitting the data to four peak lorentzian
shown in Fig.7. The average dispersion does not show any
temperature dependence with mode structure remaining same
except for superfluid near critical point.The residue
decreases with temperature for all momentum values
and the bands.The effect of temperature on residue
is stronger at small momentum values. For $U=2t$ and $10t$
we show broadening scales for only gapless band.
The broadening increases
with temperature for all of the bands and interaction
strength. The broadening scales are larger for the gapless
band on the negative frequency axis as compared to one on
the frequency axis. The broadening shows
momentum dependence and has a non monotonic dependence
on interaction strength first increasing
and then decreasing with interaction strength.

\begin{figure*}[t]
\centerline
{
~~
\includegraphics[width=17.6cm,height=6.5cm]{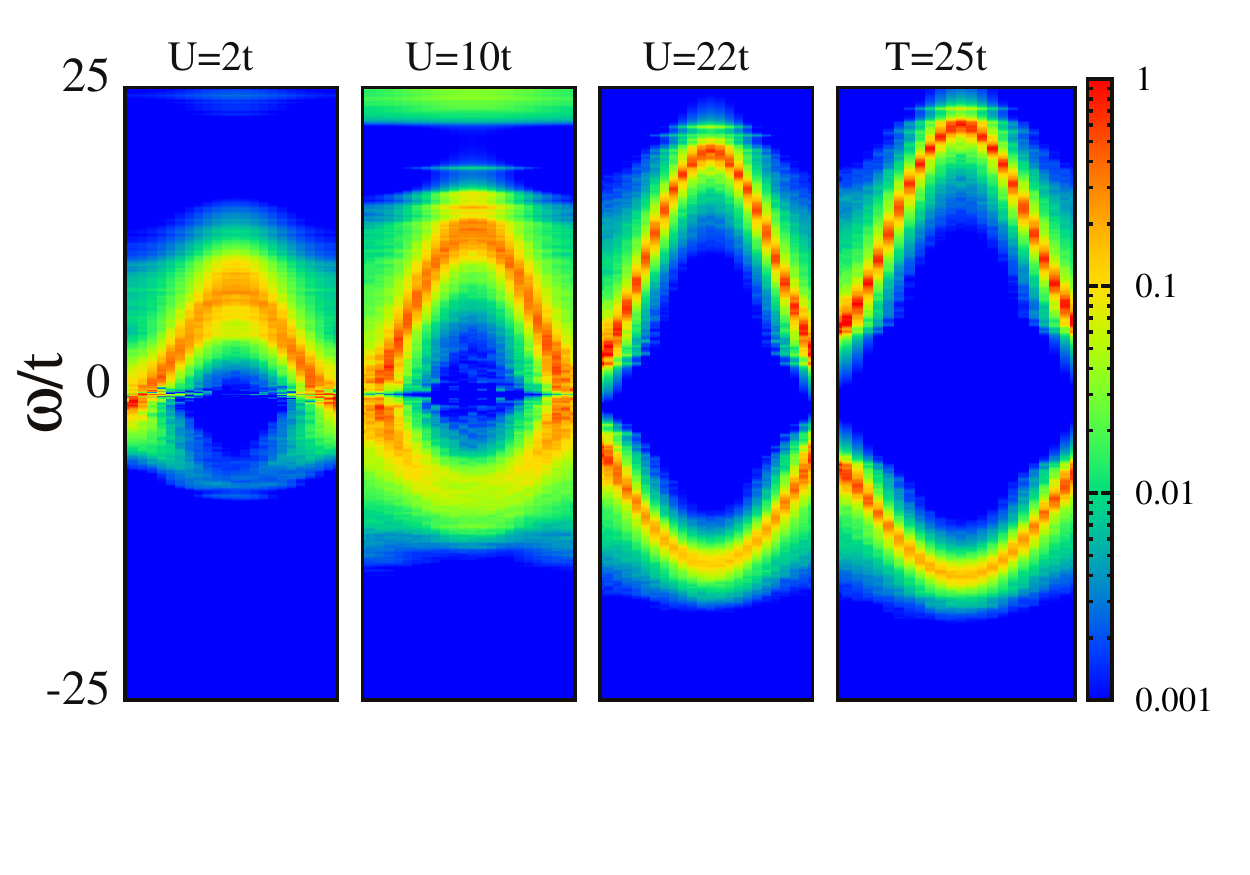}
}
\vspace{-2mm}
\centerline
{
\includegraphics[width=3.8cm,height=4.5cm]{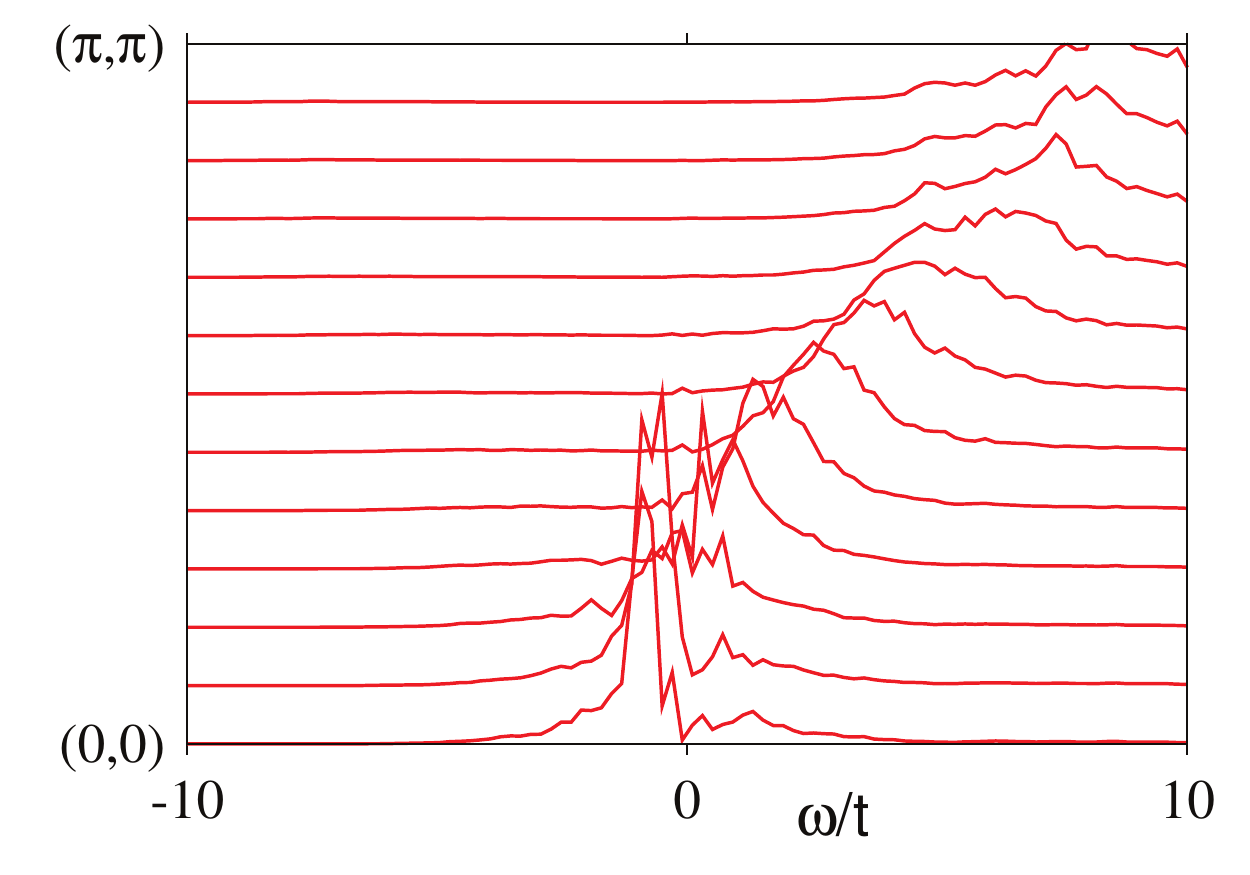}
\includegraphics[width=3.8cm,height=4.5cm]{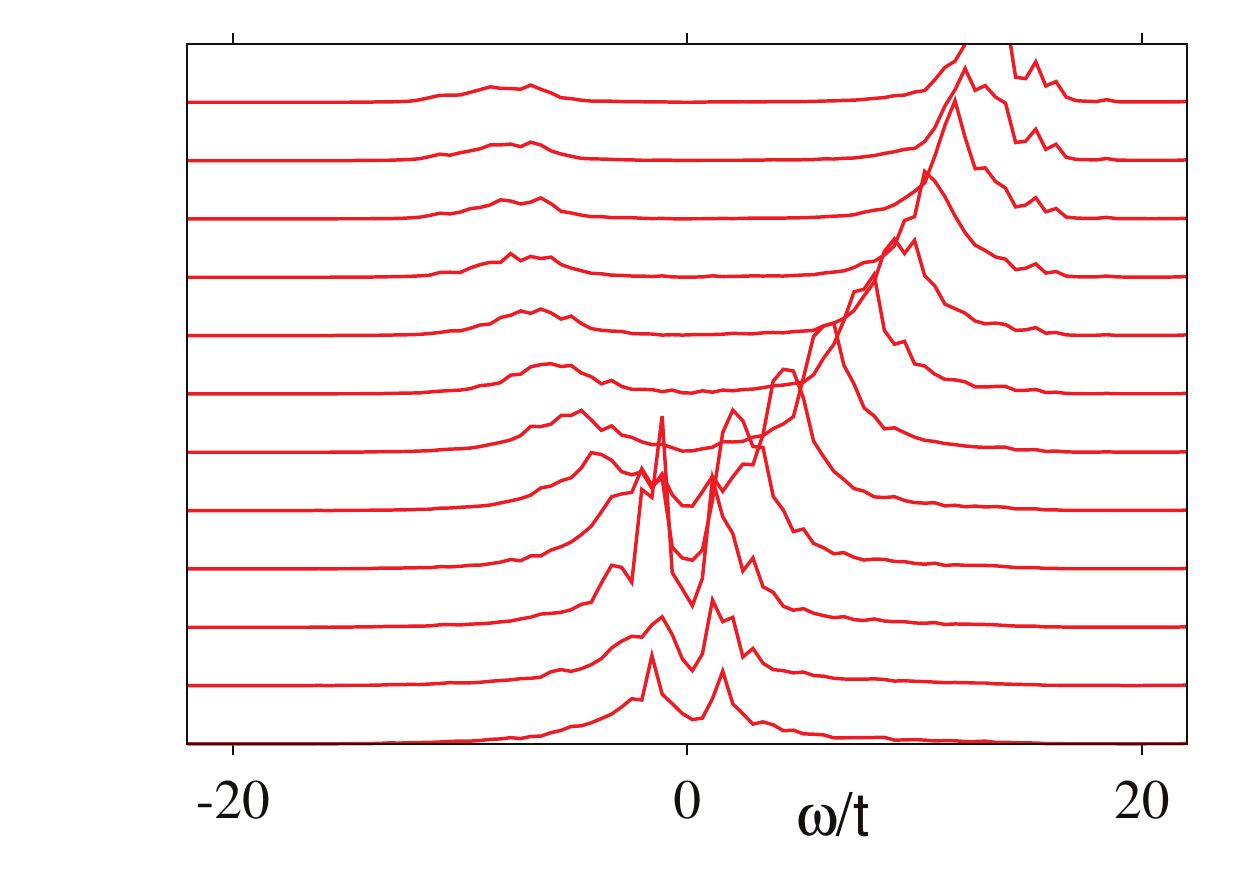}
\includegraphics[width=3.8cm,height=4.5cm]{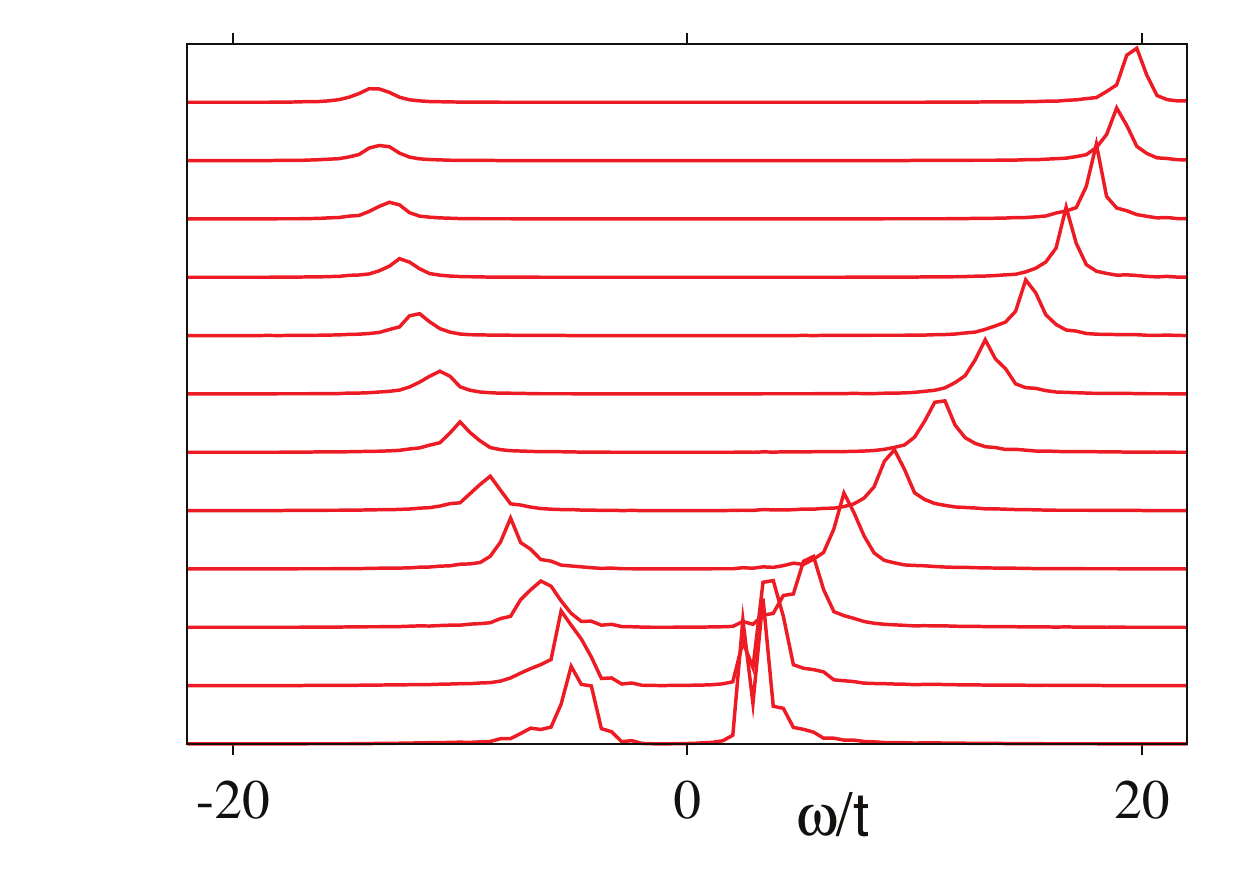}
\includegraphics[width=3.8cm,height=4.5cm]{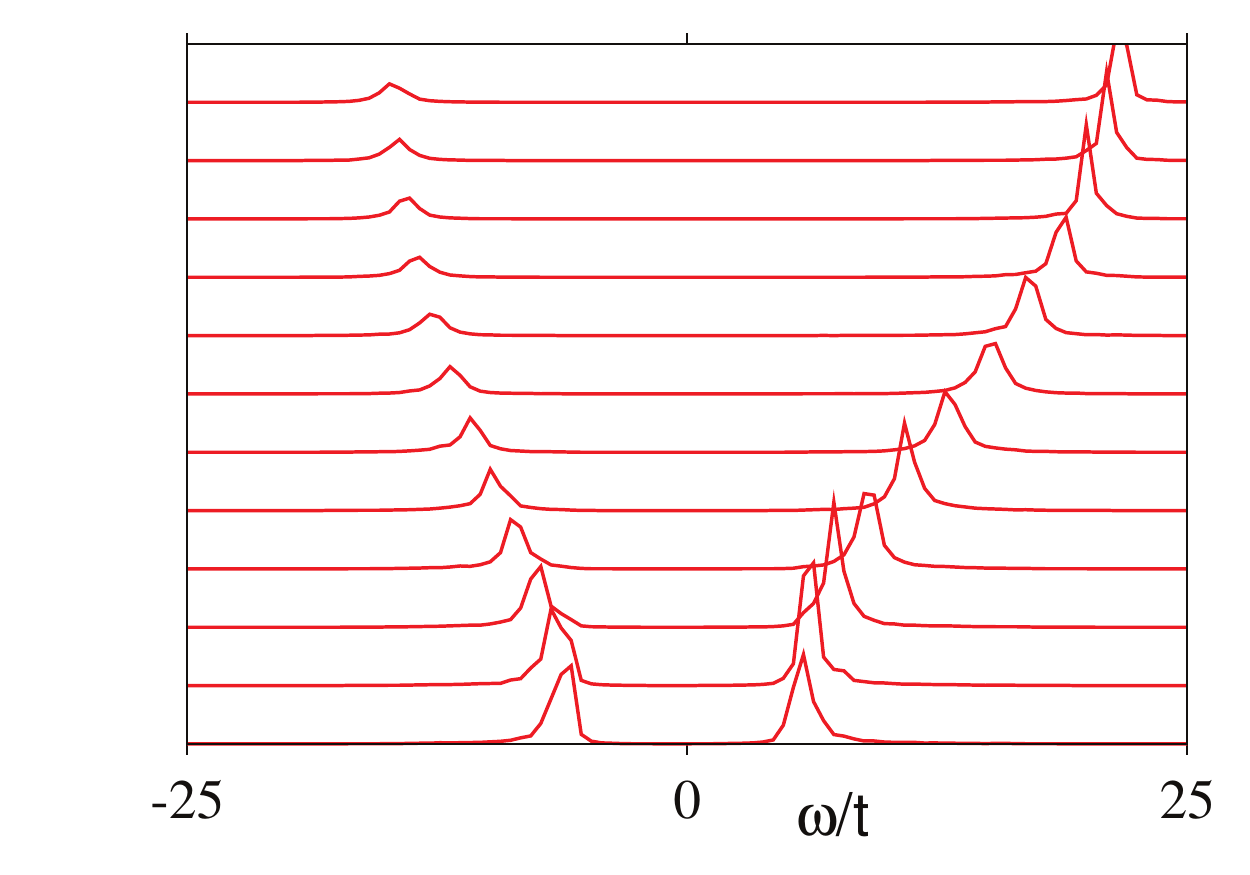}
}
\vspace{.3cm}
\centerline
{
\includegraphics[width=3.8cm,height=4.2cm]{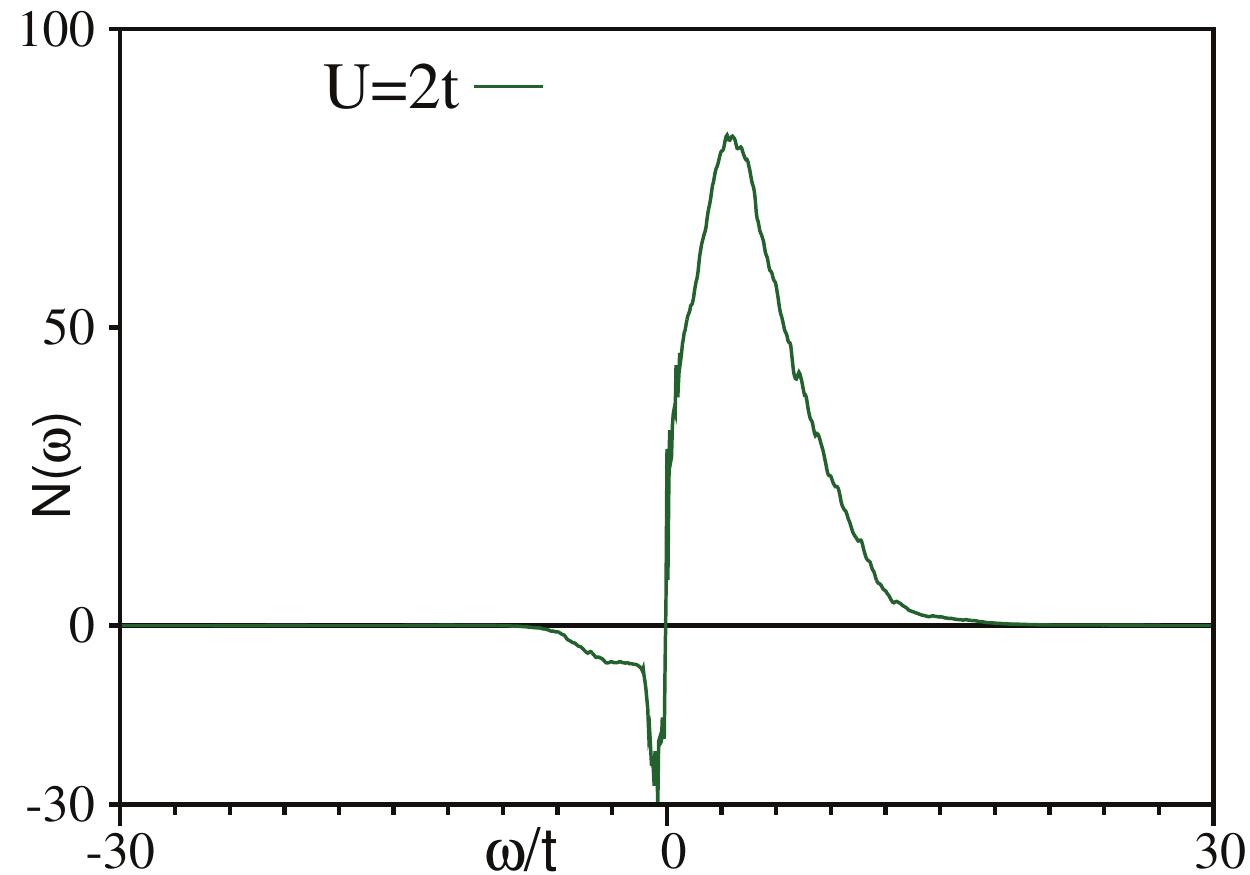}
\includegraphics[width=3.8cm,height=4.2cm]{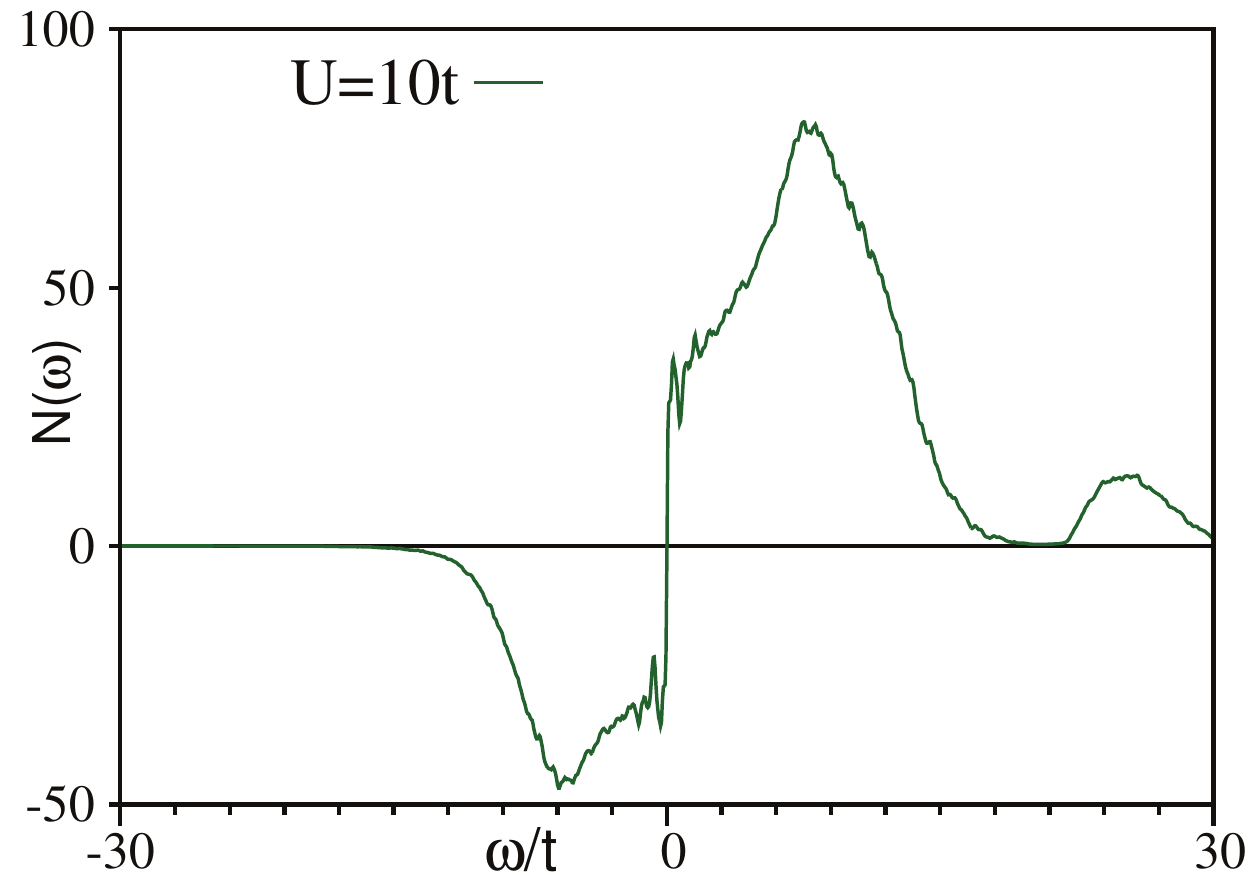}
\includegraphics[width=3.8cm,height=4.2cm]{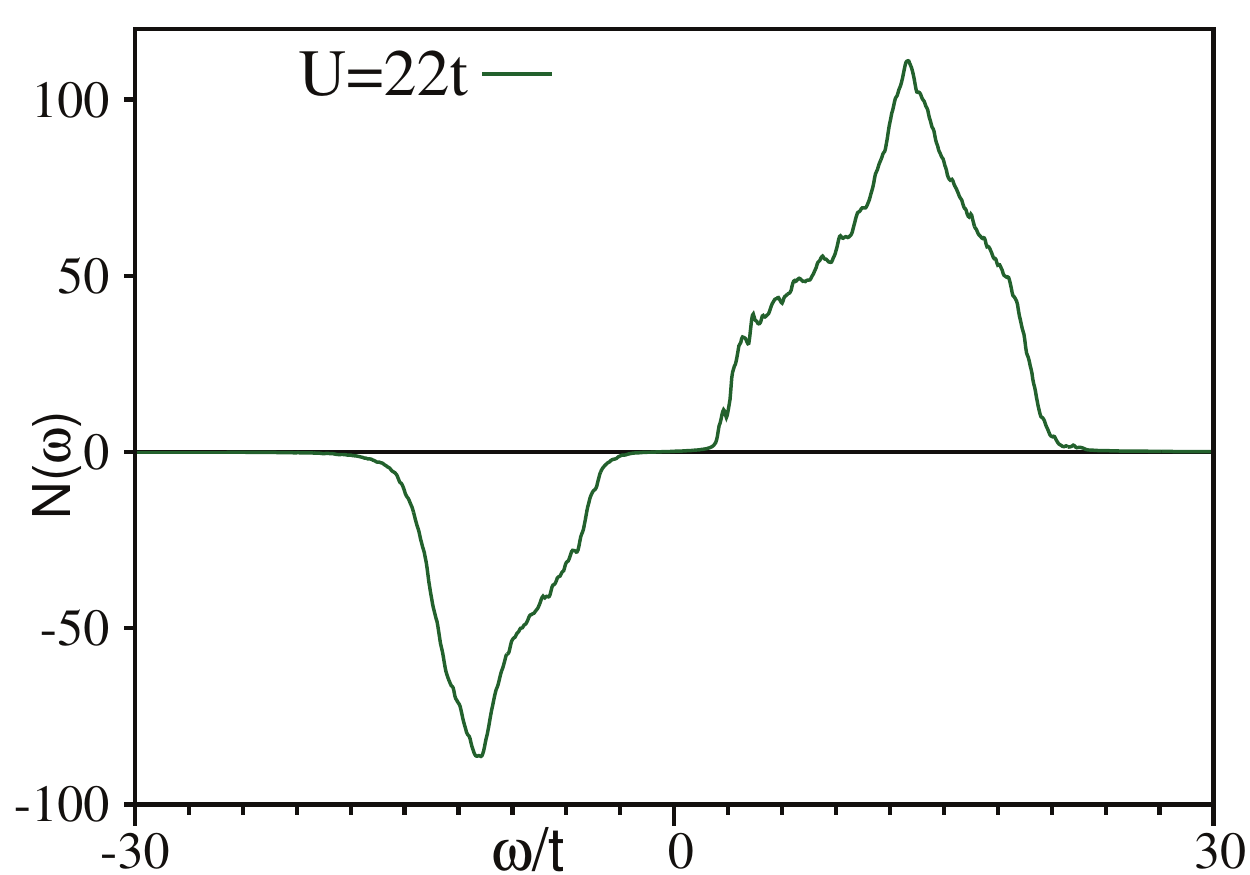}
\includegraphics[width=3.8cm,height=4.2cm]{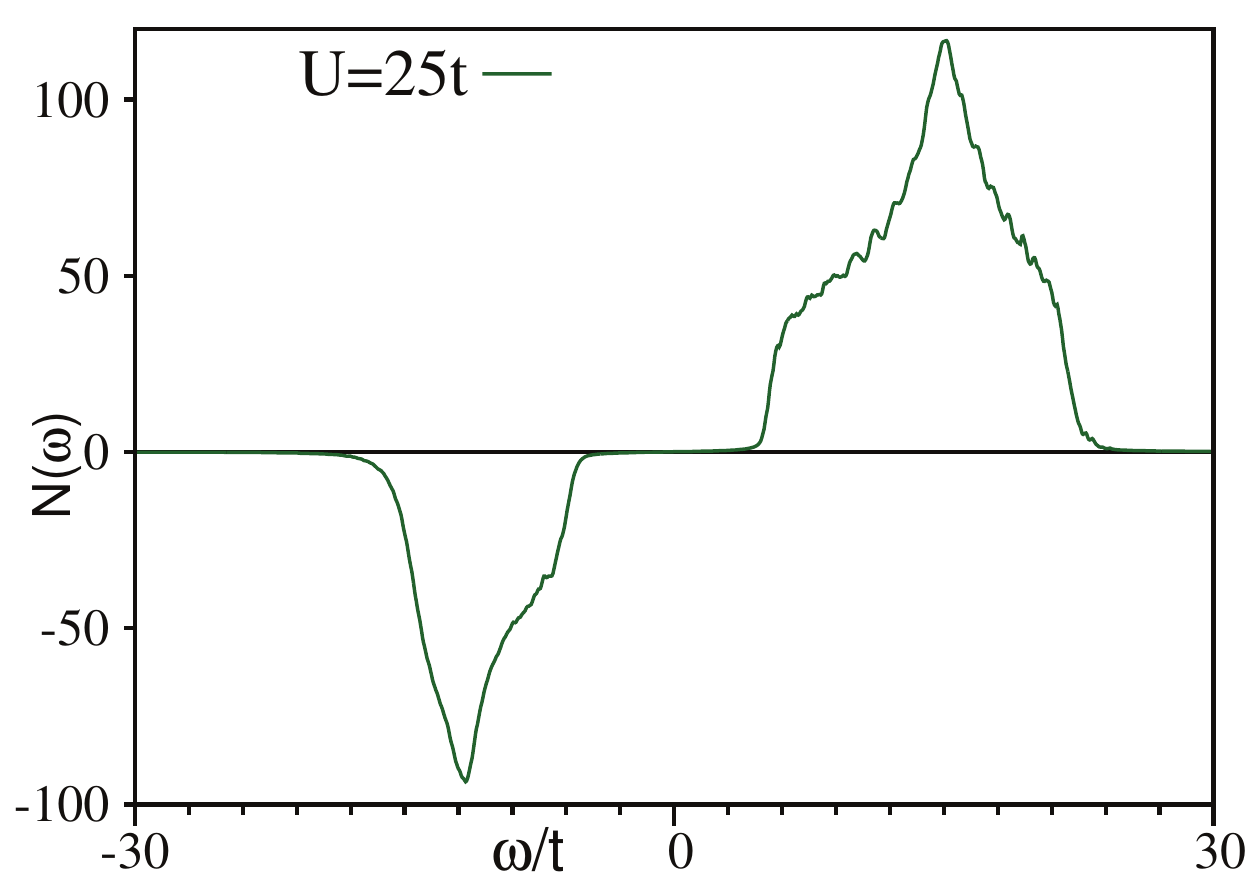}
}
\caption{The effect of increasing interaction in the
$T > T_c$  normal Bose liquid. First row:
$|A({\bf k}, \omega)|$ second row: lineshapes,
third row: N($\omega$).
From left to right $U = 2t,~10t,~22t,~25t$.}
\end{figure*}

Fig.6 shows the plot of DOS for $U=2t, 10t$ and $22t$.
For $U=2t$ the temperature dependence of
DOS is small. There is not much change in density of states
on the negative frequency side
with temperature. On the positive frequency
 the dip in the density of states
at larger $\omega$ seen at T=0t is reduced
due to mixing of amplitude and gapless
mode with rise in temperature. Overall there is
 small suppression of DOS on positive
frequency side.But near around $\omega=0$
 the rise in DOS can be seen with
temperature.
The loss of superfluidity is due to decoherence.
Similar features are seen for $U=10t$ except
that small increase in DOS peak
and shrinking of width with temperature
is visible. The similar trends in DOS can also
be seen for other $U$ values away from the
transition point. We find that even above
T$_c$ for most part of the phase diagram
DOS is not gapped around $\omega=0$.
This trend changes for
the superfluid near critical interaction strength.
At zero temperature density of states are gapless
for $U=22t$.
With rising temperature low energy peak gets suppressed
and formation of gap in DOS is visible. At high frequency
not much temperature dependence
is seen both for positive and negative frequency.

\subsection{Thermal effects in the Mott regime}

Fig.8 shows the plot of  $\vert A({\bf k}, ,\omega)\vert$ 
for $U=25t$ and $U=30t$.  The excitation spectra
consists of two modes.
Overall weight on the hole and particle
mode is seen decreasing with temperature rise.
The gap increases with temperature.
We fit the data to the two peak lorentzian
and obtain residue,mean dispersion
and broadening scales shown in Fig.10.
The average dispersion
again show not much temperature
dependence.The residue depletes with temperature
but the depletion is small.
The broadening is seen increasing
with temperature and has small momentum dependence.
The broadening scales roughly increases 
by twofold at $T=2.0t$ as compared to 
$T=0.5t$ value.
The broadening scales are similar for
the hole band as well as particle band.

Fig.9 shows the plot of  $N(\omega)$
for $U=25t$ and $U=30t$.
The effect of temperature on N($\omega$)
is sharp reduction of van-hove like peak,
increase in gap with both $U$ and $T$, but,
overall DOS retains its  feature
at all temperature.

\section{Effect of increasing interaction in the normal state}

Fig.11 shows the plot of $|A({\bf k}, \omega)|$, lineshapes
and density of states for increasing interaction at $T=3.0t.$
At $T=3.0t$ for $U=2t$ and $U=10t$ the gapless and gapped bands 
merge together and only single band is visible.The bandwidth 
increases with interaction.
Increasing interactions drives the normal bose liquid to gapped bose liquid.
The gap increases with increase in interaction strength.
This can be clearly seen from spectral map and density of states.
The asymmetry in density of states between positive and frequency 
axis reduces with increasing interaction.We fit the spectral 
data using two peak lorentzian.We find that broadening behave 
non-monotonically with increasing interaction with a maximum at 
U=4t.The mean dispersions roughly agrees with zero temperature 
dispersion except for the gapped bose liquid.The suppression
in residue is more for small interactions as compared to 
large interactions.

\section{Conclusions}

We have presented results on the single particle spectral function of 
the Bose Hubbard model at integer filling as the interaction drives 
the system from the superfluid to the Mott state. Our generalisation
of the standard mean field theory brings in all the classical
thermal fluctuations in the problem, in terms of the auxiliary field
that decouples the kinetic energy term. This, as we have
discussed elsewhere, is an excellent method for computing the
thermal transition scale in the problem. In the thermally 
fluctuating backgrounds we implement a real space
`RPA', equivalent to the `self avoiding walk' method of strong
coupling theory.  We present the momentum resolved lineshape
across the superfluid to Bose liquid thermal transition, tracking the
mean dispersion, weight and damping of the four primary
modes in the problem.
The dispersion and weight of these modes
changes with interaction but are almost temperature independent, even into
the normal state, except near critical coupling. The damping
varies roughly as $T^{\alpha} f_{\bf k}$, where $T$ is the temperature,
${\bf k}$ the momentum, $\alpha \sim 0.5$, and $f_{\bf k}$ is weakly momentum
dependent.
Near critical coupling the thermal Bose `liquid' is gapped, with progressive
widening of the gap with increasing temperature.

{\it Acknowledgement:} We acknowledge use of the High Performance
Cluster Facility at HRI. Abhishek Joshi was partially supported by
an Infosys student award. AJ acknowledges fruitful discussions with 
Sauri Bhattacharyya, Arijit Dutta and Samrat Kadge.

\section{Appendix}

We have to compute 

$G_{ij}(i\omega_n) =  Tr[e^{-\beta H} b_{j,n} 
b^{\dagger}_{i,n}]\\$

$\hspace{1.5cm}= \int 
{\cal D} \phi {\cal D} \phi^* {\cal D} \psi {\cal D} \psi^*
\text{e}^{-(S+S_b)} \phi_{j,n} \phi^{*}_{i,n}\\$

$\hspace{1.5cm}= \int 
{\cal D} \phi {\cal D} \phi^* {\cal D} \psi {\cal D} \psi^*
\text{e}^{-S}F(\phi_{j,n},\phi_{i,n}^*)\\$

$F(\phi_{j,n},\phi_{i,n}^*)=\phi_{j,n} 
\phi^{*}_{i,n}(1-S_b+\frac{S_b ^2}{2!}+...)\\$

Now,

$g_{ij}(i\omega_n) =  \int 
{\cal D} \phi {\cal D} \phi^ *
\text{e}^{-S} \phi_{j,n} \phi^{*}_{i,n}\\ $

$\hspace{1.35cm}= ~Tr[e^{-\beta H'} b_{j,n} 
b^{\dagger}_{i,n}]\delta_{ij}\\$

$g'_{ij}(i\omega_n) =  \int 
{\cal D} \phi {\cal D} \phi^*
\text{e}^{-S} \phi_{j,-n}^* \phi^{*}_{i,n}\\ $

$\hspace{1.35cm}= ~Tr[e^{-\beta H'} b^{\dagger}_{j,-n} 
b^{\dagger}_{i,n}]\delta_{ij}\\$

$
G_{ij}(i\omega_n)=\int 
{\cal D}\psi {\cal D} \psi^*(X+Y+..)\\$

$X=g_{ij}(i\omega_n)\delta_{ij};~~ Y=-\int {\cal D} \phi {\cal D} \phi^* 
\text{e}^{-S}\phi_{j,n} \phi^{*}_{i,n}S_b\\$

$
X+Y=g_{ij}(i\omega_n)\delta_{ij}
+t ~g_{ii}(i\omega_n)~g_{jj}(i\omega_n)\delta_{i-j,nn}$

$\hspace*{1.2cm}+t~g'_{ii}(i\omega_n)
g'^*_{jj}(i\omega_n)\delta_{i-j,nn}+...$

We approximate G by adding all self avoiding walk process.We 
drop corrections due to higher order loop diagrams which
are important to go beyond SPA.We  do not include corrections
due to B$_{ij}$ part of S$_b$.

The summation of self avoiding walk leads to the RPA 
corrected Green's function.The series can be more compactly 
written in a matrix notation

$$
\hat{G}\approx\int {\cal D}\psi 
{ \cal D}\psi^*e^{-\sum_i\psi^*_i\psi_i}\hat{{\cal G}}$$
$~~\text{where}~~{\cal \hat{G}}=
\frac{ \hat{g}}{(1-{\cal{T}}\hat{g})}=\frac{I}{(\hat{g}^{-1}-{\cal{T}})}$

where $\hat{g}_{ij}$ is given by $\hat{g}_{ij,n}=
\delta_{ij}~\hat{g}_{ii,n}$

$
\hat{g}_{ii,n}=\begin{bmatrix}
&\frac{Tr[e^{-\beta H_{0i}} b_{i,n} 
b^{\dagger}_{i,n}]}{Tr[e^{-\beta H_{0i}}]}  
&\frac{Tr[e^{-\beta H_{0i}} b^{\dagger}_{i,n} 
b^{\dagger}_{i,-n}]}{Tr[e^{-\beta H_{0i}}]}\\
&\frac{Tr[e^{-\beta H_{0i}} 
b_{i,n} b_{i,-n}]}{Tr[e^{-\beta H_{0i}}]} &
\frac{Tr[e^{-\beta H_{0i}} b_{i,-n} b^{\dagger}_{i,-n}]}{Tr[
e^{-\beta H_{0i}}]} 
\end{bmatrix}$

and ${\cal T}$ is the hopping matrix.
${\cal T}_{i,j}=\hat{{t}}\delta_{i-j,nn}\\$

$\hat{{t}}=\begin{bmatrix}
                  t & 0\\
                  0 & t
                 \end{bmatrix}$

We analytically continue 
$G(i\omega_n\rightarrow \omega+i\eta)$
to obtain the retarded Green function.
This is the RPA result averaged over thermal
 configurations  of the $\psi$ field.

$A({\bf k}, \omega)=-\int {\cal D}\psi 
{ \cal D\psi^*} e^{-\sum_i\psi^*_i\psi_i}  
\frac{1}{\pi}Im(\hat{v}\hat{\cal{G}}(\omega)\hat{v}\dagger)$
 where $\hat{v}$ is the row matrix ${v}_{0,i}={(e^{-i\vec{k}.\vec{r_i}}/\sqrt L,0)}$
 
 $N(\omega)=\sum_{\bf k}A({\bf k}, \omega)$

To compute the corrected G$_{RPA}$ we have to calculate atomic 
Green's function$(g)$. The atomic Green's functions are computed 
using thermal backgrounds obtained from SPA on $24 \times 24$ lattice.

\end{document}